\documentclass[11pt, a4paper,hidelinks]{article}
\usepackage[margin=1in]{geometry}
\usepackage{epsfig,amssymb,soul}
\usepackage{hyperref}
\usepackage{graphicx}
\usepackage{algpseudocode}
\usepackage{epstopdf}
\usepackage{amsfonts,amsmath}
\usepackage{bm}
\usepackage{setspace}
\usepackage{comment}
\usepackage{siunitx}
\usepackage{mathrsfs}
\usepackage{multirow}
\usepackage[hang,flushmargin,bottom]{footmisc}
\usepackage[normalem]{ulem}
\usepackage{mathtools}
\usepackage{lipsum}
\usepackage{array}
\usepackage{enumerate}
\usepackage{ifthen}
\usepackage[affil-it]{authblk}
\usepackage{cite}
\usepackage{leftidx}
\usepackage{relsize}
\usepackage{braket}
\usepackage{float}
\usepackage[ruled,vlined]{algorithm2e}

\usepackage[dvipsnames]{xcolor}
\usepackage{comment}
\usepackage{color}
\usepackage{subcaption}
\usepackage{blkarray, bigstrut}

\newcommand{\REV}[1]{{\textcolor{black}{#1}}}

\newcommand{\etal}{\textit{et al}.~}

\usepackage[right]{lineno}

\providecommand{\keywords}[1]
{\small	
\textbf{\textit{Keywords---}} #1}

\DeclareMathAlphabet\mathbfcal{OMS}{cmsy}{b}{n}
\DeclareMathAlphabet{\mathpzc}{OT1}{pzc}{m}{it}

\makeatletter
\newcommand*\bigcdot{\mathpalette\bigcdot@{.5}}
\newcommand*\bigcdot@[2]{\mathbin{\vcenter{\hbox{\scalebox{#2}{$\m@th#1\bullet$}}}}}
\algdef{SE}[SUBALG]{Indent}{EndIndent}{}{\algorithmicend\ }%
\algtext*{Indent}
\algtext*{EndIndent}

\makeatother

\title{Time transient Simulations via Finite Element Network Analysis: Theoretical Formulation and Numerical Validation}

\author[a]{Mehdi Jokar${}^*$}
\author[a]{Siddharth Nair}
\author[a]{Fabio Semperlotti\thanks{To whom correspondence should be addressed. Email: fsemperl@purdue.edu, mjokar@purdue.edu}}
\affil[a]{School of Mechanical Engineering, Ray W. Herrick Laboratories, Purdue University, West Lafayette, IN 47907}
\date{}

\begin{document}

\maketitle

\begin{abstract}
This paper extends the finite element network analysis (FENA) to include a dynamic time-transient formulation.
FENA was initially formulated in the context of the linear static analysis of 1D and 2D elastic structures. By introducing the concept of \textit{super finite network element}, this paper provides the necessary foundation to extend FENA to linear time-transient simulations for both homogeneous and inhomogeneous domains. The concept of neural network concatenation, originally formulated to combine networks representative of different structural components in space, is extended to the time domain. Network concatenation in time enables training neural network models based on data available in a limited time frame and then using the trained networks to simulate the system evolution beyond the initial time window characteristic of the training data set. 
The proposed methodology is validated by applying FENA to the transient simulation of one-dimensional structural elements (such as rods and beams) and by comparing the results with either analytical or finite element solutions. Results confirm that FENA accurately predicts the dynamic response of the physical system and, while introducing an error on the order of 1\% (compared to analytical or computational solutions of the governing differential equations), it is capable of delivering extreme computational efficiency.  
\end{abstract}

\section{Introduction}\label{sec1:Intro}

The concept of finite element network analysis (FENA), introduced by Jokar and Semperlotti in~\cite{jokar2022two, jokar2021finite}, is a deep-learning-based simulation framework for structural systems; although the basic concept is theoretically extendable to simulate the evolution of any physical system for which surrogate models can be defined. Combining the modularity and flexibility of FEA with the computational speed of trained neural networks (NN), FENA proposes a fundamentally novel approach to the efficient and accurate machine learning-based simulation of physical systems with an emphasis on overcoming some critical limitations typical of neural network-based simulation approaches. 
From this perspective, one of the most remarkable properties of FENA is its ability to build models of systems by interconnecting pre-trained (surrogate) neural network models representative of selected components. These pre-trained networks are available in an existing database and do not require any \textit{ad hoc} training after being connected to other networks. This property is made possible by introducing the concepts of \textit{network concatenation} combined with a proper \textit{library of network elements}.
In FENA, NN-based surrogate models (dubbed \textit{network elements}) of various physical systems are pre-developed (i.e. designed and trained) and stored in a database, that is the \textit{library of network elements}. Networks within the library are typically trained to model specific classes of problems. As an example, in our previous work~\cite{jokar2022two} we developed libraries of slender beams and thin plates subject to generally distributed transverse loads and boundary conditions. Many libraries can be developed and used, depending on the physical systems to be simulated in FENA. 
Once the library of surrogate elements is available, the concept of network concatenation allows the creation of a composite system model, that is a model formed by some combination of selected (pre-trained) network elements from the library. Once again, following the example of thin plates and assuming the availability of a library of beam and plate elements, a stiffened panel could be modeled in FENA by assembling (i.e. concatenating) the corresponding network finite element models~\cite{jokar2022two}; no training would be required after the model is built, in fact training is not required at all because the individual networks were already pre-trained. This property marks a substantial departure from other network-based simulation methods for physical systems that instead require specific training for every specific configuration, input, and boundary condition.

In~\cite{jokar2021finite, jokar2022two}, we applied FENA to the static simulation of structures. Specifically, we developed a library of network elements to simulate general classes of fundamental structural elements with various geometric and material properties, various input loads, and boundary conditions. 
The extension of FENA to the dynamic range and, more specifically, to time transient analysis requires some significant modifications in order to account for the challenges introduced by time dependence in systems responses.
This study makes two main contributions toward achieving this goal.
First, we present the concept of transient \textit{super finite network element}, which provides a generalized fundamental architecture for the development of transient network elements. 
We provide a general methodology to design super finite network element subcomponents that are capable of reproducing transient solutions to initial boundary value problems with homogeneous and inhomogeneous domain properties. We also present the design guidelines that are crucial to the training convergence of the network elements for relatively more sophisticated domains, such as domains with inhomogeneities. 
\REV{The second contribution is the introduction of generalization (extrapolation) capability by extending the concept of network concatenation to the time domain.} Network concatenation extends the prediction (time) window of a network element by feeding the input back into an analogous network. 
This procedure is numerically shown to enable simulations over extensive time ranges with networks trained on time windows that are only a fraction of the simulated time range. \\

\noindent \textbf{Brief review of deep-learning-based scientific computing}\\

Recent advances in computational resources and cloud computing, combined with the revolution in deep learning algorithms have led the scientific community to explore the potential of these methods hold to support numerical simulations. From a general perspective, deep neural networks provide two important advantages when considered as a possible substitute for classical numerical computations. 
The first property refers to the ability of a neural network to learn the behavior of virtually any system. This property is guaranteed by the universal approximation theorem~\cite{Goodfellow-et-al-2016} that states that a neural network can approximate any complex nonlinear mapping provided sufficient training data, training iterations, and proper network architecture. 
The second property refers to the comparatively low computational cost of trained neural networks. The small computational cost originates from the intrinsic nature of a neural network that, once trained to mimic a physical system's response, does not require solving a system of differential equations to evaluate the response.
The price to pay is the computational cost associated with the generation of the training datasets and with the training of the models. 

The recent surge of interest in applications of deep learning techniques has resulted in abundant publications dedicated to neural network-based scientific computation. Examples include nonlinear systems~\cite{9158400, baiges2020finite, parish2016paradigm}, fluid mechanics~\cite{mao2020physics, jin2021nsfnets}, nano-optics~\cite{chen2020physics}, and physical systems with fractional governing equations~\cite{pang2019fpinns, patnaik2022role}.  
From the perspective of network architecture design and training, the available literature could be divided into two main categories: 1) data-driven and 2) physics-constrained approaches. We briefly review each approach and discuss its benefits and challenges. Note that, given the large volume of studies produced on this topic in recent years, the following should not be seen as a comprehensive review but rather as a short overview of the concept. 

Data-driven approaches rely on deep neural networks trained exclusively based on available data. Typically, this class of methods requires large databases in order to achieve successful training.
Data-driven approaches leverage an important advantage of deep learning algorithms, that is the ability to detect features in large volume of data. This allows the development of NN-based surrogate models of physical systems~\cite{wu2021design, brevis2021machine}, deep learning-assisted numerical methods~\cite{mishra2018machine, meister2020deep}, and NN-based inverse problem solvers~\cite{patnaik2022distillation, patnaik2022variable}.
The diversity of physical systems modeled with neural networks demonstrates the flexibility of deep learning-based approaches to model physical phenomena governed by various types of PDEs.

Despite the unparalleled success in reducing computational cost, network-based surrogate models face two critical limitations. 
First, they are not generalizable. Many studies, including those cited above, have demonstrated that network-based surrogate models can achieve high prediction accuracy across training and test datasets. However, there is no guarantee that the networks will exhibit the same level of accuracy when tested with out-of-distribution sample data. These studies do not provide clear directions (similar to the process of convergence analysis in FEA) for assessing the reliability and accuracy of the predictions where no ground truth solution is available. Hence, the level of accuracy when a trained network is used to simulate unseen data (i.e. sample cases that do not belong to the test or training datasets) is somewhat unknown.
Secondly, another critical limitation of network-based surrogate models is their need for problem-specific (i.e. \textit{ad hoc}) training. 
The surrogate NN-based models are \textit{customized} for a specific system and input-output configuration. It follows that any change in the problem setup (such as in domain properties or boundary and initial conditions) requires a new set of training dataset generation followed by network training. This problem, also known as the issue of \textit{customized models},  is critical because the training phase is the most time-consuming operation in neural network applications, as it can be extremely computationally expensive 
even with the most advanced available hardware~\cite{ling2016reynolds}. Hence, although the computational cost of prediction for trained neural networks is minimal and is very unlikely to exceed its classical numerical method counterpart, the cost of retraining may rapidly annihilate the computational benefit of network-based modeling.

On the other side, physics-driven techniques implement the available knowledge about physical systems in NN-based scientific computations. The main goals of knowledge implementation are to improve the accuracy of the network predictions and to reduce the dependency on training data. Physical laws can be incorporated into neural networks in three different ways: 1) revised training loss function~\cite{raissi2019physics, kharazmi2021hp} which is also referred to as physics-informed neural networks (PINN), 2) network architecture design~\REV{\cite{VLASSIS2020113299, jin2020sympnets, hernandez2021structure}}, and 3) network initialization~\cite{readprocess, wu2022physics}; the first type is undoubtedly the most popular approach.

Recall that neural networks are trained by minimizing a function called loss function. In the first physics implementation approach, the available expert knowledge can be incorporated into the network by revising the training loss function, which measures the consistency of predictions with the governing equations or constraints of the system. This became possible by the concept of backpropagation~\cite{Goodfellow-et-al-2016}, which allows analytical calculation of the derivatives of the network output with respect to its inputs. 
Following this concept, Raissi~\etal successfully solved different types of PDEs with PINNs without using any training data samples~\cite{raissi2019physics}. During the training, randomly distributed collocation points in the input parameters space were used to validate the consistency of network predictions with physical laws. Since the introduction of the concept of PINNs, the idea of PINNs has been successfully applied to a wide range of domains, such as to solve problems in computational fluid mechanics~\cite{mao2020physics, jin2021nsfnets}, nano-optics~\cite{chen2020physics}, acoustics~\cite{nair2024multiple}, heat transfer~\cite{cai2021physics}, identification of governing PDE of physical systems~\cite{raissi2019physics, doan2019physics}, and inverse problems~\cite{nair2023grids, jin2020physics}. 

The key benefit of physics-constrained models is their potential to learn to replicate a system without requiring training data~\cite{raissi2019physics, cai2021physics}, making them suitable for situations where training data are unavailable. 
The neural networks obtained with this approach can be envisioned as a transfer function that directly maps a system input to its response without numerically solving the governing equations. 
Despite this computational advantage, the main limitation is the need to redesign and retrain the network every time the system's physical parameters change; because the networks are customized and specifically trained for an assumed combination of domain properties and boundary conditions. This aspect is very detrimental from the perspective of developing a computational framework for the simulation of physical systems when considering the high computational cost of training. 

Further, network models are trained using an optimization procedure with a (highly likely) nonconvex loss function~\cite{pantidis2023error}. Training may converge to a local minimum where the training loss (calculated on a set of randomly selected collocation points) is minimal. 
Recall that one of the challenges in training neural networks is that they tend to converge the low-frequency components of the network outputs rather than capturing all the details included in the high-frequency components of the output, which can reduce the training convergence and, consequently, the prediction accuracy~\cite{cui2022efficient}. 
This is a critical aspect when we consider that PINNs training may not converge when applied to systems with stiff governing PDEs~\cite{wang2021understanding}, in which the PDE solution contains sharp variations, typical of the response seen in transient simulations and wave propagation problems. We highlight that the other two approaches for physics implementation also result in customized models and training convergence challenges for transient simulations, hence bearing the same limitations discussed above. 

\REV{The above discussion reveals that the majority of deep learning-based approaches are limited to a fixed set of material properties, boundary and initial conditions, and geometric properties. Hence, any change to the system condition necessitates regeneration of the training dataset followed by network training. This limitation is very important considering the computational cost associated with network training and is the main barrier to the development and widespread adoption of deep learning-based simulations. Recently, researchers suggested solutions to reduce the computational cost associated with training~\cite{jokar2021finite, raissi2019physics, wang2021train}. However, pursuing solutions to this critical limitation is still an open area of study. 
As discussed in the section above, the concept of FENA is designed to present some very unique capabilities that can address the critical issue of customized models by introducing the concepts of the library of network elements and network concatenation.} 

Building upon previous work that focused on the static simulation of structures~\cite{jokar2021finite, jokar2022two}, this paper extends the FENA concept to include transient simulations. To accomplish this, we present modifications to the instructions provided in~\cite{jokar2021finite, jokar2022two} that enable the modeling of transient effects. 
The remainder of this paper is organized as follows. In \S~\ref{sec: SFNE description}, we first present the general architecture for the super finite network element (SFNE) of FENA, followed by the introduction of the concatenation algorithm for the assembly of SFNEs in the time domain.  
\S~\ref{sec: rod}~and~\ref{sec: beam} are dedicated to the application of FENA (and the proposed concepts) to the transient simulation of initial boundary value problems. Specifically, \S~\ref{sec: rod} demonstrates different realizations of the SFNE applied to the transient simulation of homogeneous rods, and \S~\ref{sec: beam} covers the application of the proposed concepts to the transient response of inhomogeneous beams.
In each section, we compare the developed network elements with analytical or numerical solutions to validate the performance of the proposed methodologies in the transient simulation of structures. 

\section{Extension of FENA to dynamics: Fundamentals}\label{sec: fundamentals}

This section presents general guidelines for extending FENA to transient simulations. 
In this regard, we will introduce two key concepts: 1) super finite network element (SFNE), and 2) network concatenation in the time domain.
SFNE is a generalization of the finite network element introduced in~\cite{jokar2022two}, which allows capturing the dynamic behavior of a physical system in transient simulations.
The network element concatenation in time domain refers to the process of interconnecting multiple SFNEs in order to simulate the system far beyond the time range each SFNE was trained for. In this section, we present the details of the aforementioned concepts, while in \S~\ref{sec: rod} and \S~\ref{sec: beam} we discuss the details of applying them to transient structural simulations.  

\subsection{Super finite network element}\label{sec: SFNE description}

As discussed in \S~\ref{sec1:Intro}, in its original formulation, FENA~\cite{jokar2022two, jokar2021finite} was limited to static problems. Specifically, in~\cite{jokar2022two, jokar2021finite}, the architecture of finite network elements (FNE) was designed to handle sequences of properties in the spatial domain. The concept was targeted to the realization of network models of inhomogeneous systems (either because of material, loading, or boundary conditions).
In order to enable the application of the FNE concept to transient analyses, the general architecture of the FNEs must be substantially redesigned.
To this end, we propose a new neural network architecture, referred to as \textit{super finite network element}, which is based on bidirectional recurrent neural networks (BRNN)~\cite{Goodfellow-et-al-2016}. The proposed architecture of SFNE allows the incorporation of temporal information into the neural network, hence enabling capturing transient effects in the inputs and outputs of the network.    

\begin{figure}[!h]
	\centering
	\includegraphics[width=.99\linewidth]{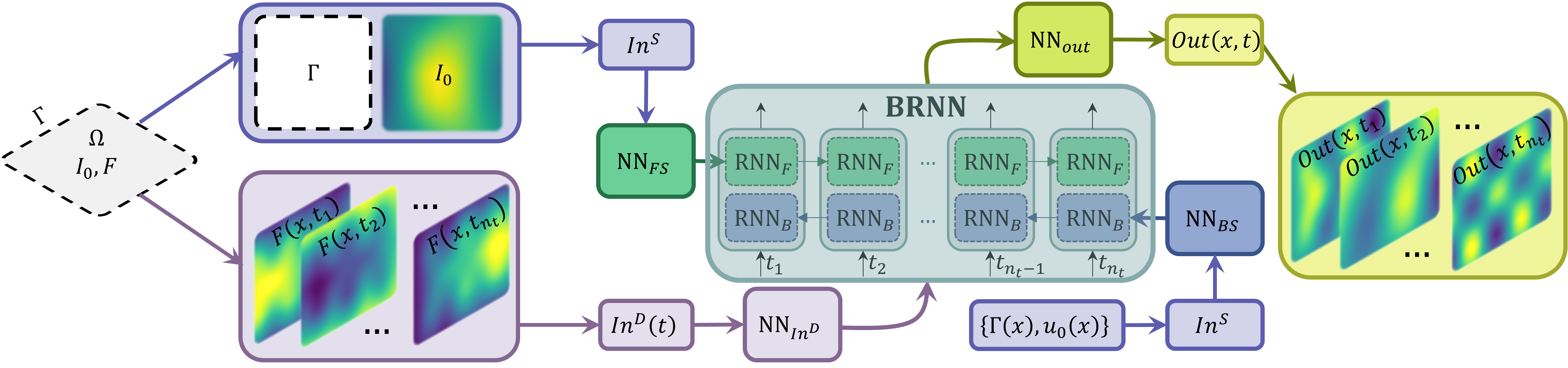}
	\caption{Schematic showing the architecture of transient super finite network element of FENA. Consider domain $\Omega$ that is subject to the initial condition $I_0$, boundary condition $\Gamma$, and input $F$. $t$ and $x$ represent time and space domains, respectively. Using the general architecture of SFNE, surrogate models of $\Omega$ can be built to predict its response $Out(x,t)$ to $I_0, \Gamma, F$ at $\{t_1, ..., t_{n_t}\}$.}\label{fig: super_NE_Arch}
\end{figure}

\REV{Figure~\ref{fig: super_NE_Arch} presents the general architecture of SFNE that serves as the foundation to extend FENA to the transient domain. The overall network architecture and functionalities builds on three major components: a) the \textit{input}, b) the \textit{neural network core (BRNN)}, and c) the \textit{output}. While their specific use will be discussed in \S~\ref{sec: rod}~and~\ref{sec: beam}, we will discuss their general role in detail in the following sections.}


\subsubsection{Input}

\REV{The inputs to the network model are provided via two branches labeled $In^S$ and $In^D(t)$ (see Figure~\ref{fig: super_NE_Arch}). All the physical properties, boundary conditions, and initial conditions that remain constant in time (static) are provided through $In^S$, while $In^D(t)$ consists of a sequence of all dynamic properties, such as applied external force inputs, at discrete time steps $\{t_1, t_2, ...., t_{n_t}\}$. $In^D = \{In^D_1, In^D_2, ..., In^D_{n_t}\}$ consists of a sequence of all dynamic properties and inputs at the time steps $\{t_1, t_2, ..., t_{n_t}\}$ such as applied external inputs.
Through $In^D$, we highlight that the SFNE (potentially) has the ability to handle time-evolving properties, such as memory effects~\cite{patnaik2020applications} or time-varying boundary conditions~\cite{hintermann1989evolution}. However, in this study, we focus only on domain parameters and conditions that are constant over time. Before passing the inputs to the network core, $In^S$ is passed through a block of neural network layers ($NN_{FS}$ and $NN_{BS}$). Similarly, $In^D(t)$ is passed to another block of neural network layers named $NN_{In^D}$. The inputs are processed through these neural network blocks to provide additional learning capacity and enhancing their expressivity~\cite{lin2018generalization}. Recall that expressivity refers to the ability of deep neural networks to approximate some functions that cannot be approximated with shallow nets (having the same number of neurons). Hence, the addition of NN blocks enhances the ability of SFNEs to approximate complex patterns and sharp variations typical of transient responses. Note that the network blocks ($NN_j$ where $j=\{FS, BS, In^D \}$) introduced here are not independent neural networks, rather they are stacked neural layers that function with the neural network core and output.}
        
Particularly in the case of $In^S$, it is passed to $\text{NN}_{FS}$ and $\text{NN}_{BS}$ to initialize the forward and backward initial hidden states of the BRNN. 
Recall that the hidden states in a BRNN are the internal states of the network that store information about the history of the input data and the current state of the network. Hidden states allow the network to remember information from previous inputs and use it in the prediction of future time steps. While the initial states of a recurrent neural network are typically set to be zero~\cite{Goodfellow-et-al-2016}, we offer a new strategy to initialize recurrent units. We leverage hidden states initialization to inform the network of the system properties that remain constant with time. In other terms, we adjust the initial states of the forward and backward recurrent units of the BRNN block according to the system (constant) parameters and conditions. This allows the model to consider the effect of relevant system information that is not explicitly part of the dynamic input sequence data ($In^D$), thereby enhancing the model's predictive capability and versatility.
 
Further, we highlight that $In^S$ should only be used when varying parameters are considered as input to the SFNE. For example, for a SFNE that models a class of problems having the same material properties, it is not necessary to include the properties in $In^S$. The reason is that such properties remain the same in the entire training dataset. Hence, the network implicitly learns this without being explicitly informed about the numeric values of those parameters. Note that if all constant parameters are the same across all training data samples, we set $In^S = 0$ (see, for instance, the case in \S~\ref{sec: BV rod}). On the other hand, in the case where SFNE is used to model systems with arbitrary inhomogeneity, $In^S$ is used to inform the network about the distribution of inhomogeneous material properties within the computational domain (see the problem discussed in \S~\ref{sec: beam}). Different forms of $In^S$ will be presented in the example cases discussed in the following sections.

\subsubsection{Neural network core (BRNN)}

\REV{The neural network core consists of a bidirectional recurrent neural network (BRNN) architecture. The BRNN is composed of two sets of recurrent neural network (RNN) cells, each one processing the sequence of input data in opposite directions, that is forward ($RNN_F$) and backward ($RNN_B$) in time along with the static properties (including boundary and initial conditions). While the data processed through $NN_{FS}$ and $NN_{In^D}$ are used as inputs to $RNN_F$, the $RNN_{B}$ uses $NN_{BS}$ and $NN_{In^D}$ as its inputs. Here, $NN_{FS}$ and $NN_{BS}$ represent the initial hidden state inputs that carry information between time steps, and $NN_{In^D}$ represents the dynamic input. Finally, the outputs of the $RNN_F$ and $RNN_B$ cells are combined to calculate the global output at each time step. It is important to emphasize the significance of the sequence of RNN cells, where each instance of the RNN cell corresponds to a time step. In BRNN, these $RNN_F-RNN_B$ cells serve as a sequence learning model by constantly passing key features from previous (forward) or later (backward) time steps. This continuous passage of critical data forms the foundation of sequence learning, aimed at comprehensively capturing the system dynamics through sequential data ~\cite{Goodfellow-et-al-2016}. }

In static analysis~\cite{jokar2022two, jokar2021finite}, the use of BRNNs was well justified due to the presence of boundary conditions. By incorporating information from both ends of the sequence, the BRNNs were able to effectively propagate this information (i.e. the nature of the boundary conditions) to different spatial locations within the domain and accurately model the behavior of the system. This aspect was also critical to be able to achieve network elements that can simulate different boundary conditions without further re-training.
In dynamic analysis (time domain simulations), the use of BRNN is still critical to deal with boundary conditions while they play a different role when it comes to processing time sequences. The causality of the time response would simply need a RNN (single direction) architecture. \REV{However, BRNNs deliver a superior ability to learn and capture complex dependencies within data sequences compared to RNNs; we found that this was valuable for accurate predictive modeling. Further, the increased contextual understanding of BRNNs can result in faster learning and reduce the dependence on extensive datasets when compared to RNNs (especially in scenarios with complex dynamics). In contrast, RNNs solely rely on past information, which limits their contextual understanding and may require a larger amount of data and more training epochs to achieve comparable accuracy~\cite{cui2020stacked, wang2022deep, BRNN-Schuster1997, althelaya2018evaluation}. The reader is remanded to~\cite{jokar2022two} (see \S~2) and \cite{jokar2021finite} (see \S~2 and \S~3) for a detailed discussion on the internal architecture of BRNNs and the role of the bidirectional flow of information in the simulation of structures via FENA.
Regarding the computational cost, although BRNNs are slightly slower than unidirectional RNNs due to the doubling of the recurrent cells, we found this trade-off acceptable in light of the improved prediction accuracy. It should be noted that from the perspective of computational efficiency, trained BRNNs still have significantly lower computational time compared to FEA or analytical solutions (see the discussions in \S~\ref{sec: computational costs} and \S~\ref{sec: IV beam}). The slight increase in the computational cost is a minor price to pay to achieve higher accuracy due to the bidirectional flow of information.}

\REV{Unlike traditional recurrent neural networks (RNNs) that possess a "causal" architecture (due to the unidirectional information flow), BRNN architectures are considered "non-causal" as they incorporate information from future time instances \cite{turek2020approximating}. However, while the architecture itself is non-causal, the causality of the final output is not affected since only the hidden layers (not the final outputs) are shared between different time instances (see Figure~\ref{fig: super_NE_Arch} in the manuscript). In other terms, while the RNN and BRNN can be classified as causal and non-causal, based on their architectures, both can be successfully used for causal predictions \cite{yoon2017rnn, horvath2022granger}. Indeed, it is insightful to observe that all networks for sequence learning (including RNNs and BRNNs) learn the problem's dynamics based on the complete sequence knowledge prior to model training. Therefore, the causality of the model is enforced through the prior causal training data used for sequence learning. Additionally, the superior prediction accuracy of BRNNs over RNNs underscores the efficiency of their non-causal architecture in effectively optimizing network weights to capture the underlying dynamics accurately.}

\subsubsection{Output}

\REV{The inputs when passed through the neural network core approximates the outputs of the network $Out(x,\textbf{t})=\{Out(x,t_1), Out(x,t_1), ..., Out(x,t_{n_t})\}$ (see Figure~\ref{fig: super_NE_Arch}). More specifically, the network predicts a sequence of responses for all the input time steps in $In^D$, starting from the first time step ($t_1$) to the last time step ($t_{n_t}$) through the $RNN_F-RNN_B$ cells. Similar to the network blocks used in the input, the outputs from the $RNN_F-RNN_B$ cells are passed through another block of neural network layers represented $NN_{out}$ to evaluate the final output $Out(x, \textbf{t})$.}

Note that $x$ may be either a single point output location or a group of $n_x$ numbers of nodes distributed within the spatial domain ($x=\{x_1, x_2, x_{n_x}\}$). The number of nodes is user-defined and is defined based on a set of locations (or even a single point) at which the response is sought. If a complete spatial domain response is needed, then $x$ represents an evenly spaced grid of nodes (see \S~\ref{sec: BV rod}~and~\ref{sec: on-demand location}). 
Note that the output dimension $n_x$ scales linearly with the number of collocation points, which can potentially pose a challenge when dealing with complex geometries. However, several strategies can be followed to overcome this limitation. First, the modularity of FENA and the concept of network concatenation can be leveraged to handle complex geometries. We can decompose complex geometries into simpler subdomains, each represented by a separate network element from the library. These elements can then be interconnected to form the overall system model, allowing us to manage complex geometries potentially without increasing the output dimension~\cite{jokar2022two}. 
Second, as discussed in \S~\ref{sec: on-demand location}, SFNEs can be trained to receive the prediction location (i.e. the collocation points) as an input parameter provided via $In^s$. This approach decouples the network output prediction from the size of $n_x$, which provides another solution to the scaling issue. 
Further, CNNs and autoencoders can be used to handle systems with large $n_x$. CNNs are known for their ability to efficiently handle high-dimensional data~\cite{Goodfellow-et-al-2016}. CNN can be used in the $\text{NN}_{In^D}$ and $\text{NN}_{out}$ blocks to extract spatial correlations in the solution and reduce the number of parameters (SFNEs size) required to predict the system's response (see \S~\ref{sec: beam} for an example of using CNNs in the SFNE $\text{NN}$ blocks). Autoencoders can be particularly useful for dimensionality reduction. They are a type of neural network designed to learn an efficient, compressed representation of data, which can be especially beneficial when dealing with high-dimensional output spaces. An example application of autoencoders to high-dimensional data can be found in~\cite{wu2022physics}. \\


\REV{In summary, a generalized SFNE network architecture with separate network blocks to provide flexibility and describe different system configurations for transient simulations is introduced. These supporting network blocks may consist of different combinations of neural network layers tailored for specific applications (see \S~\ref{sec: rod}~and~\ref{sec: beam} for details of this procedure).} The SFNE is trained to predict $Out(x,t)$ at $t=\{t_1, t_2, ..., t_{n_t}\}$ corresponding to a given set of $In^D$ and $In^S$. We highlight that $In^S$ and $In^D$ can contain various conditions and external inputs. Hence, trained SFNEs cover a category of problems without requiring retraining.
Note that SFNEs will be trained over a predefined prediction window ($t\in[t_1, t_{n_t}]$), and their accuracy is reduced for $t>t_{n_t}$. To alleviate this limitation, in the following we introduce the concept of network concatenation, which enables accurate predictions far beyond the prediction window that the SFNE has seen during its training ($t \gg t_{n_t}$). In the remainder of the paper, the term \textit{network element} refers to the super finite network element introduced in this section. 

\subsection{Network concatenation: extending the prediction window of SFNEs}\label{sec: concat description}

As discussed above, SFNEs are trained to predict the response for $n_t$ time steps ($t\in[t_1, t_{n_t}]$). It follows that the SFNEs prediction period (window) is limited to $t\in[t_1, t_{n_t}]$, hence predictions made beyond this period may have lower accuracy.
This is an important limitation that can prevent the application of FENA to transient simulations because it limits the network elements to the prediction window seen during the training. In this section, we discuss how this limitation can be addressed by extending the idea of concatenation~\cite{jokar2021finite, jokar2022two} to time domain problems. It will be later shown that the proposed concatenation algorithm can be used to assemble the network elements in the time domain and to perform simulations for long periods of time, beyond the training prediction window of SFNEs. 

The initial conditions of a physical system can be provided to SFNEs through $In^s$ (see \S~\ref{sec: SFNE description}). The SFNE is then trained to predict a system response up to $n_t$ time steps for a given set of initial conditions and external input ($In^D$).
In other words, the SFNE is an initial value solver for a general set of initial conditions. To extend the network prediction window beyond the training window of $n_t$ time steps, we implement an iterative approach that repeatedly uses this initial value solver (the trained SFNE). Specifically, we first predict the response by using the network for the first $n_t$ time steps. Then, we use the prediction at the time step $n_t$ as the initial condition to simulate the system response for the next time steps $n_t$, that is, $n_t+1, ..., 2n_t$. The proposed procedure, which effectively implements a \textit{network concatenation} in the time domain, can be applied any arbitrary number of times in order to calculate the response of a physical system at any arbitrary time step, far beyond the prediction window the network was trained for. \REV{In summary, the concatenation in time is essentially a sequence of time evolutions of the system (in chunks equal or smaller to the time window used for training) under different initial conditions. If the FENA modules are well trained for initial conditions, the dynamics of the system will be fully learned by the network, and changing initial conditions will not affect the ability of the network to evaluate the response.}

The concatenation procedure requires two important precautions to maintain accuracy over time. Because while the network can be very accurate when tested on the test datasets (e.g., see the results in \S~\ref{sec: initial BV rod}), even a minimal error in the predictions (including the error in the last time step) can propagate and build up following the successive concatenations of SFNEs in time. This mechanism might drastically reduce the accuracy even after a few concatenation steps. In this case, the decrease in accuracy is not associated with the SFNE itself but it rather originates from the small error in the initialization of the network (inaccurate $In^S$) that gets repeated at every concatenation step. To control the error accumulation, we propose the following adjustments; both aim at enhancing the accuracy of the initialization process at every step of the concatenation:

\begin{itemize}
    \item \ul{Model Ensemble}: The first revision of the above concatenation strategy is based on the concept of \textit{model ensemble}. Ensemble modeling refers to the process of combining the predictions of multiple models to boost prediction performance and to improve overall accuracy. Model ensemble techniques are based on the observation that a group of networks results in enhanced prediction quality~\cite{ensembleofnets} compared to the case where a single model is used for prediction. The gain in accuracy is due to the relatively lower prediction accuracy of a model being compensated with the higher accuracy offered by the rest of the models in the ensemble.    
    Following a model bagging approach~\cite{breiman1996bagging}, an ensemble of models can be used to enhance the overall accuracy of the predictions, hence the initial conditions used in the subsequent concatenation steps. 
    According to this strategy, $In^S$ is defined at each concatenation step based on the average response predicted by the ensemble of network elements. Using an ensemble of models appreciably enhances the accuracy of initialization at each concatenation step, hence increasing the prediction accuracy at the following time steps and significantly decelerating the error accumulation during concatenation. \newline
    An important aspect of using a model ensemble approach is the determination of the number of models needed in the ensemble. The ensemble size can be determined by gradually increasing the number of models and assessing the prediction error in a test dataset. While prediction accuracy improves significantly for the first few models added to the ensemble, it eventually reaches a point where minimal improvement is seen upon increasing the ensemble size any further (e.g., see Figure~\ref{fig: rod_concat_sample}a). 
    \item \ul{Cut-off threshold}: As we will see in the prediction results of SFNEs (e.g. Figure~\ref{fig: rod_concat_sample}b), the SFNE's relative error is not uniform over all time steps. In fact, the prediction accuracy of SFNEs is slightly reduced as we approach the $n_t^{th}$ time step.
    This is a phenomenon typical of recurrent neural networks that tend to gradually lose accuracy when progressing in the prediction sequence, especially in larger sequences~\cite{Goodfellow-et-al-2016}. To control the error accumulation, we propose to identify and define a cut-off threshold step ($t_c < t_{n_t}$). This can be done at the network performance evaluation step where ground truth is available (test dataset), and prediction error can be calculated. The threshold step is the time step at which the relative error begins to rapidly increase.
    At each step of concatenation, we should only use the predictions up to the cut-off threshold and disregard the rest. The cut-off threshold can be easily identified from the model ensemble relative error versus the time-step graph (for example, see the results in \S~\ref{sec: IV_BV_rod time concat}). 
    We also note that using the response only up to $t_c$ requires additional concatenation steps to obtain the response up to a desired time step (compared to the case where no sequence cutting is performed). For example, if the goal is to obtain the response up to $8n_t$ and $c= 0.8n_t$, 10 concatenation steps ($10\times 0.8n_t = 8n_t$) are needed to predict the response over the desired period. However, without cutting the response at $t_c$, concatenation can be performed in 8 steps. Nevertheless, neural networks prediction is extremely fast, hence performing a few more network evaluations barely affects the computational time. On the other hand, enhancing $In^S$ accuracy significantly improves network concatenation accuracy. 
\end{itemize}

Algorithm~\ref{Algorithm:concat} summarizes the above-discussed steps for network elements concatenation. Before we proceed further, we clarify a few aspects associated with the suggested concatenation algorithm. 

First, it should be noted that in FENA a single network element (or ensemble of elements) is trained to predict the response of a class of problems rather than of a specific configuration (e.g., initial condition). This means that only one set of trained SFNEs is needed to simulate a system and also to implement the proposed concatenation algorithm, regardless of the complexity of the system. In this sense, FENA draws some parallel with the traditional finite element analysis, where element types (e.g., beams, plates, solids) are general classes of elements that can be used to model any complex structure resulting from their combination. 
It follows that during concatenation, the trained SFNE (or ensemble of SFNEs) is executed multiple times to predict the response of the system at each concatenation step. This eliminates the need for multiple copies of the same weights of the network elements, significantly reducing the memory footprint of the approach.

Second, we highlight that the primary purpose of the proposed network element concatenation algorithm is to provide a robust, scalable, and computationally efficient approach for systems where the dynamics of the system remains relatively consistent within the problem space defined by the network elements. The developed framework allows us to effectively and accurately model a large variety of systems by selecting and combining the pre-trained network elements that are trained with data from a limited time frame. This is an outstanding property, even in linear systems, because it alleviates the need for recurrent training of network elements if the user is interested in time periods extending beyond what was seen by the network elements during the training phase (see the results presented in \S~\ref{sec: IV_BV_rod time concat}). Further, concatenation is useful in cases where (training) data in the desired simulation time frame is unavailable, as network element retraining is not possible due to the unavailability of training data.

\begin{algorithm}[!htbp] 
\caption{Network element concatenation. }\label{Algorithm:concat}
\Indm building model ensemble:\\
\Indp   \While{ensemble accuracy improves}{sample from training dataset\\ train and add a SFNE to the ensemble}
\Indm determine the ensemble $t_c$ from prediction error versus time step graph.\\
simulation:\\
\Indp \While{the desired simulation period ($t_{max}$) is not reached}{initialize network element via $In^S$\\ predict system response \\ save the predictions up to $t_c$ \\ use the predictions at $t_c$ to form $In^S$ for the next step of network concatenation}
\end{algorithm}

\section{Transient simulations via FENA: Application to homogeneous systems}\label{sec: rod}

This section presents the application of FENA to the transient simulations of a one-dimensional homogeneous structure. Specifically, this section aims at 1) providing general instructions and considerations when FENA is applied for transient simulation, and 2) showcasing the ability and performance of FENA for transient analysis. Note that while solving the 1D wave equation within a homogeneous domain is well established from a mathematical and numerical perspective, its accurate solution via neural networks is a more challenging task especially due to the presence of oscillating patterns. This problem is made even more challenging if a general solution is sought; this is the case of the present study where the network is required to deal with multiple initial and boundary conditions without any further training. Recall that most of the existing network-based approaches to the simulation of physical systems require retraining for every set of boundary and initial conditions (and, in some cases, applied inputs). Hence, the development of NN-based modeling approaches that can deal with, at least, a complete class of problems, even if for relatively simple 1D geometries, is still of great importance. \REV{Further, in this specific study, we focused on the development and implementation of FENA to analyze temporal variations on a fixed geometric mesh. In other words, this work allows considering general loadings and boundary conditions, as well as different time-dependent dynamics but on structures at fixed spatial coordinates (discretized nodes).} In the following section, we show the capability of FENA when applied to inhomogeneous materials and geometries and discuss in detail the issues associated with such category of problems. 

In the following, we first focus on the domain solution of the time domain 1D wave equation, which can be seen as representative of the longitudinal response of a 1D rod. The following three cases are considered: 1) the rod is subject to a boundary force while initial conditions are zero (see \S~\ref{sec: BV rod}); 2) the rod is subject to nonzero boundary and initial conditions~(\S~\ref{sec: initial BV rod}); and 3) the rod is subject to distributed loads with nonzero initial conditions (\S~\ref{sec: dist load rod}).

\subsection{Case~1: Transient response to boundary excitation}\label{sec: BV rod}
\subsubsection{Problem Statement}\label{sec: BV Rod Problem Statement}
\begin{figure}[!h]
	\centering
	\includegraphics[width=0.8\linewidth]{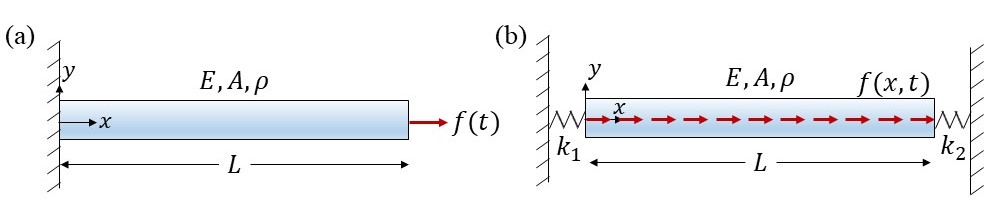}
	\caption{\REV{Schematic illustrations of the homogeneous rods for (a) \textbf{Case-1 and Case-2}: having constant properties ($E$, $A$, and $\rho$) and an axial boundary load $f(t)$ applied at $x=L$. Note that cases-1 and 2 differ in the initial conditions. (b) \textbf{Case-3}: having constant properties ($E$, $A$, and $\rho$), stiffness boundary conditions ($k_1$ and $k_2$), and a uniformly distributed harmonic load $f(x,t)$.}}\label{fig: rods_schematic}
\end{figure}

Consider a thin cantilever rod aligned along the $x$-axis and having length $L$, uniform cross-section $A$, Young's modulus $E$, and density $\rho$ as shown in Figure~\ref{fig: rods_schematic}a. The rod is fixed at $x=0$ and is free at $x=L$. The rod is subject to an axial boundary load $f(t)$ applied at $x=L$. The governing equation of the system is given by: 
\begin{equation}\label{eq: BV_rod}
\begin{gathered}
    E A \frac{\partial^2 u}{\partial x^2} = \rho A \frac{\partial^2 u}{\partial t^2}\\
\begin{aligned}
    u(x,0)&= u_0(x) & \dot{u}(x,0)= \frac{\partial u(x,0)}{\partial t} &= \dot{u}_0(x) \\
    u(0,t)&= 0
     & EA \frac{\partial u(L,t)}{\partial x}&= f(t)\\
\end{aligned}
\end{gathered}
\end{equation}

\noindent where $u$ is the rod displacement. The goal is to develop a SFNE that receives $f(t)$ as input and calculates the dynamic response of the system considered as initially at rest (i.e. $u_0(x)=0$ and $\dot{u}_0(x) = 0$). Also, for this case, we assume that $E$, $A$, and $\rho$ are constant and do not vary across the training samples, and the only input in Case~1 is the boundary load $f(t)$. 

\subsubsection{SFNE architecture}\label{sec: Case 1 SFNE Arch}

Table~\ref{table: BV_rod_netArch} presents the details of the network element architecture developed for the rod discussed in \S~\ref{sec: BV Rod Problem Statement}. The core BRNN block has 100 long-short-term memory (LSTM)~\cite{Gers1999} cells (50 LSTM cells in each forward and backward direction). 
We chose LSTM specifically to address the vanishing gradient problem commonly encountered in simple RNNs. Recall that LSTM units employ a gating mechanism that helps propagating information over long sequences without significant loss of relevant information, which is an important feature considering the nature of our simulations~\cite{Goodfellow-et-al-2016}. 

The activation function $ES$ in Table~\ref{table: BV_rod_netArch} stands for E-Swish activation function~\cite{E-swish-alcaide2018} defined by: 
\begin{equation}\label{eq: eswihs-activation}
	ES(x) = \frac{\beta x}{1+e^{-x}}
\end{equation}

The parameter $\beta$ of the E-Swish activation functions is initialized to $1$ at the beginning of the network training phase. $\beta$ is then learned during the network training step along with the weights of the network element. 

The network dynamic input ($In^D$) is the time history of the axial boundary load $f(t)$, provided to the network as a sequence of length $n_t$ defined by $In^D = \left\{f(t_i), f(t_2) ..., f(t_{n_t})\right\}$. 
Since all the properties of the problem domain are constant and the initial conditions are zero in this case, we set $In^S =0$ (see the discussions in \S~\ref{sec: SFNE description}). 

Further, recall that LSTM has two parameters (internal states) that transfer information among time steps~\cite{Gers1999} and require initialization at the first time step. The states are 1) cell state, which is the long-term memory of the network, and 2) hidden state, which is the working (short-term) memory of the LSTM unit. As discussed in \S~\ref{sec: SFNE description}, SFNE uses separate NN blocks to initialize each internal state independently from the rest. Specifically, $\text{NN}_{FS}$ (see Figure~\ref{fig: super_NE_Arch}) consists of two MLP network blocks, $\text{NN}_{FCS}$ and $\text{NN}_{FHS}$, which initialize the forward LSTM units cells and the hidden states, respectively. Similarly, $\text{NN}_{BS}$ is composed of $\text{NN}_{BCS}$ and $\text{NN}_{BHS}$ that initialize the backward cell and hidden states, respectively. Note that the size of the last layer in each of these MLPs must be equal to the number of LSTM units in the BRNN block. For example, 50 LSTM units in the forward direction require a vector of length 50 to initialize its hidden states. 

In this problem, we assume that the network element predicts the response of the entire spatial domain and at all the time steps $\{t_1, t_2, ..., t_{n_t}\}$. Hence, the network output at each time step is a vector representing the rod deformation on $n_x$ equally spaced nodes. Note that the SFNE output layer size is always the same as $n_x$ and it has a linear activation function. The numerical values of $n_x$ and $n_t$ should be determined based on the problem conditions and parameters and will be discussed in the following section (\S~\ref{sec: Training_BV_Problem}). 
We highlight that while the network in~\cite{jokar2021finite} was used as a basis to determine the present network element architecture, \REV{we adopted a structured grid-like approach to find its architectural hyperparameters (layer size, number of layers, and combinations of activation functions) via a trial-and-error procedure.} 

\begin{table}[!h]
	\caption{Summary of network element architecture used for Case~1. (AF: activation function). The values within the size vector show the number of neurons within a hidden layer, and the length of the size vector shows the total number of layers. The AF vector is the activation function of each layer. For example, size = \{10,10\} and AF = \{tanh, ES\} represent a MLP with 2 hidden layers with 10 neurons in each layer, where the first layer has $tanh$ activation function and the second layer has E-Swish activation function. 
	\label{table: BV_rod_netArch}}
	\centering
	\resizebox{.7\columnwidth}{!}{%
    \begin{tabular}{|c|c|c|c|c|c|}
        \hline
         \multirow{3}{*}{$\text{NN}_{FHS}$} & type&  MLP & \multirow{3}{*}{$\text{NN}_{BHS}$}& type& MLP \\
         & size& $\{20,50\}$ & & size & $\{20,50\}$\\
         & AF  & \{$tanh, ES$\} & & AF & \{$tanh, ES$\}\\
         \hline
        \multirow{3}{*}{$\text{NN}_{FCS}$} & type&  MLP &  \multirow{3}{*}{$\text{NN}_{BCS}$}& type& MLP \\
         & size& $\{20,50\}$ & & size & $\{20,50\}$\\
         & AF  & \{$tanh, ES$\} & & AF & \{$tanh, ES$\}\\
         \hline
        \multirow{3}{*}{$\text{NN}_{In^D}$} & type& MLP & \multirow{3}{*}{$\text{NN}_{{out}}$}& type& MLP\\
         & size& $\{20,60,60,60\}$& &size & $\{50,60,60,60,60,60,60\}$\\
         & AF& \{$tanh, ES, ES, ES$\} & & AF & \{$ES, ES, ES, ES, ES, ES, ES$\}\\
         \hline
    \end{tabular}
	}
\end{table}
\subsubsection{Network training and prediction results}\label{sec: Training_BV_Problem}
The dataset used to train and evaluate the network element consisted in $6\times10^4$ data samples.
Table~\ref{table: rod Params} summarizes the numerical values of the rod material and geometric properties. 
The boundary load was assumed to be harmonic and defined by $f(t)= f_0 \sin(\omega_0 t)$, where $f_0 =1~N$.
Note that the functional form of the input load can be general and does not need to be harmonic. We chose a harmonic boundary load for this network element because it allows obtaining an analytical solution of Eq.~\ref{eq: BV_rod}, and provides an efficient method to generate the required training datasets. This selected loading condition also allows full transparency on conceptual aspects of the proposed methodology without clouding the description of the approach with unnecessary considerations on the numerical methods needed to support more generalized loading patterns. 
However, recall that the goal of FENA is not to have a separate network element for every possible loading condition. The proposed SFNE architecture and the FENA approach are independent of any specific loading condition. 
This choice for the loading condition does not mean that the network elements are limited to a specific loading pattern (such as the harmonic patterns mentioned above) but rather that they can be trained to cover a range of possible patterns. Indeed, the ability to model various loading conditions is achieved by generating a diverse training dataset that covers a wide range of loading conditions. Once trained, the network elements will be able to predict the response to unseen loading conditions within the same class of problems.
Furthermore, given that the problem is linear with respect to $f_0$, the system response corresponding to $f_0 \neq 1$ is obtained by scaling the network element solution predicted for $f_0 =1$. The analytical solution of the governing equations is given by (see~\cite{doyle2020wave} for the details of the analytical solution):

\begin{equation}\label{eq: sol_BV_rod}
\begin{gathered}
u(x,t) = \sum\limits_{r=1}^{\infty} U_i(x)\eta_i(t) \\
U_i(x) = \sqrt{\frac{2}{\rho A L}}\sin\left(\frac{\omega_i x}{c}\right),~
\eta_i(t) = \frac{U_i(L)}{\omega_i}\int\limits_0^t {f(t-\tau)\sin\left(\omega_i\tau\right)d\tau}\\
\omega_i = \frac{(2r-1)\pi}{2L}c,~c=\sqrt{\frac{E}{\rho}} 
\end{gathered}
\end{equation}

\noindent for a harmonic boundary load defined as $f(t)= f_0 \sin(\omega_0 t)$, $\eta_i$ is given by:

\begin{equation}\label{eq: etar_BV_rod}
\eta_i(t) = \frac{(-1)^{(r+1)}f_0}{\omega_i}\sqrt{\frac{2}{\rho A L}}
\frac{\omega_0 \sin(\omega_i t) - \omega_i \sin (\omega_0 t)}{\omega_0^2-\omega_i^2}
\end{equation}

\begin{table}[!h]
 \caption{Rod dimensions and material properties. The upper and lower bands for $\omega_0$ are selected based on the first and the fourth natural frequencies of the rod. The choice of the bands was arbitrary and any other combinations for upper and lower bands could be selected. \label{table: rod Params}}
  \centering
  \resizebox{.55\columnwidth}{!}{%
  \begin{tabular}{|c|c|c|c|c|}
    \hline
    $L$ & $A$ & $\rho$ & $E$& $f(t)$\\
    \hline
    \multirow{2}{*}{$1~m$} & \multirow{2}{*}{$10^{-4}~m^2$}& \multirow{2}{*}{$9000~kg/m^3$}&
    \multirow{2}{*}{$10^7~Pa$}&
    $f_0 \sin(\omega_0 t)$\\
     & & & & $\omega_0 \in [52.5, 366.5]~rad/s$\\
    \hline
  \end{tabular}
  }
\end{table}

\REV{Each data sample of the dataset is generated by randomly selecting $\omega_0$ from the range mentioned above} and calculating the response for $t\in [0,500]~ms$ with a time step of $1~ms$, hence $n_t = 500$. In this problem, we set $n_x=51$ (i.e. the network output is a vector of size 51) that describes the response of the system in 51 evenly spaced grid of nodes on the rod. Note that the numerical values for the prediction time step and $n_x$ are based on the problem parameters and were chosen to avoid aliasing the output within the frequency range mentioned above (Table~\ref{table: rod Params}). Also, the series solution (Eq.~\ref{eq: sol_BV_rod}) was cut-off at $r_{max} = 200$.

A network element may be viewed as a nonlinear mapping between its inputs and outputs. The mapping is controlled by a set of trainable parameters $\mathbf{\Theta}$. For the network element of this section, the trainable parameters include the network layers weights and biases plus the E-Swish activation functions parameters ($\beta$). $\mathbf{\Theta}$ is calculated during the procedure known as network training, which minimizes a loss function $\mathcal{L}$ based on the mean squared error (MSE) of the predictions and the true values obtained from the training dataset. The network element training can be expressed as:

\begin{equation}\label{eq: network training}
    \mathbf{\Theta^*} = \min_{\mathbf{\Theta}}~ \mathcal{L}\left(\mathcal{N}(In^D, In^S;\mathbf{\Theta}), {Out}^{true}(x,t)\right)
\end{equation}

\noindent where $\mathbf{\Theta^*}$ is the trained network parameters, $\mathcal{N}$ is the network element, and ${Out}^{true}(x,t)$ is the true output value corresponding to the input ${In^S}$ and $In^D$, provided by the training dataset. 

The generated dataset was divided into training and test sets. 85\% of the samples were randomly selected to train the network (training dataset), and the remaining 15\% were used to assess the predictions of the network (test dataset). Note that we used the same train test split ratio ($85\%-15\%$) for the rest of the SFNEs discussed in this paper.
We highlight that the test set in our study served a dual purpose. It was used as a validation set, in the sense that it helped us monitor the SFNE's performance during training to prevent overfitting (but not for the purpose of early stopping or hyperparameters tuning). On the other hand, it was also used as a test dataset to assess the performance of trained network elements.

We used Python PyTorch to build the SFNE and trained it with ADAM~\cite{adam} optimization algorithm. The training was performed for a total of 1000 epochs. The learning rate (LR) was initially set at $.001$ and divided by a factor of two every 300 epochs (implemented using PyTorch learning rate scheduler). Table~\ref{tab: training_Params} summarizes the hyperparameters used for training the network elements. Note that we used the same training algorithm and training hyperparameters for all the network elements presented in \S~\ref{sec: rod}.
The training was carried out using a NVIDIA Tesla V100 GPU installed on an HPC cluster node equipped with a Xeon Gold 5218R CPU with 40 cores at 2.10GHz base frequency and 192 GB of memory.
Note that although a larger batch size could potentially make better use of the V100's computational capacity and accelerate the training process, we used a batch size of 64, considering the performance and generalizability of our models. The decision to use a relatively small batch size in training the models was driven by multiple considerations. First, the focus was not solely on computational efficiency but also on the quality of the trained models. Larger batch sizes may converge faster in terms of computational time but tend to result in poorer generalization performance. This is because larger batch sizes can lead to a loss of sensitivity to subtle patterns in the data, which is crucial for capturing transient simulations and dynamic behavior. 
Second, larger batch sizes could introduce a bias towards samples with larger amplitudes, potentially neglecting those with smaller amplitudes. 
Third, smaller batch sizes provide a regularizing effect and can improve the model's generalization performance because the gradient (of the network elements' weights calculated via backpropagation) estimate is less accurate with fewer samples within the batch. Hence, smaller batch sizes introduce more noise during training, preventing overfitting~\cite{keskar2016large}. 
Lastly, while larger batch sizes may fill up the GPU memory more effectively, leading to faster computation per epoch, they can also require more epochs to converge. Hence, it is not guaranteed that larger batch sizes always result in shorter overall training times. In conclusion, we found that batch size 64 provided a good balance between computational efficiency and model performance.

Here we developed only one network element because in this section we are mainly focused on presenting the steps needed to develop the basic architecture. However, for the subsequent sections, we developed an ensemble of SFNEs to enhance prediction accuracy and also for network concatenation (\S~\ref{sec: IV_BV_rod time concat}).

\begin{table}[!h]
 \caption{Network element training hyperparameters. The learning rate scheduler step indicates the number of epochs after which the LR scheduler divides the learning rate by the division factor. \label{tab: training_Params}}
  \centering
  \resizebox{.6\columnwidth}{!}{%
  \begin{tabular}{|l|c|l|c|}
    \hline
    total training epochs & 1000& initial LR& 0.001\\
    \hline
    batch size & 64 & LR scheduler step& 300\\
    \hline
    training algorithm & ADAM &LR scheduler division factor & 2 \\
    \hline
  \end{tabular}
  }
\end{table}

\begin{figure}[!h]
	\centering
	\includegraphics[width=.8\linewidth]{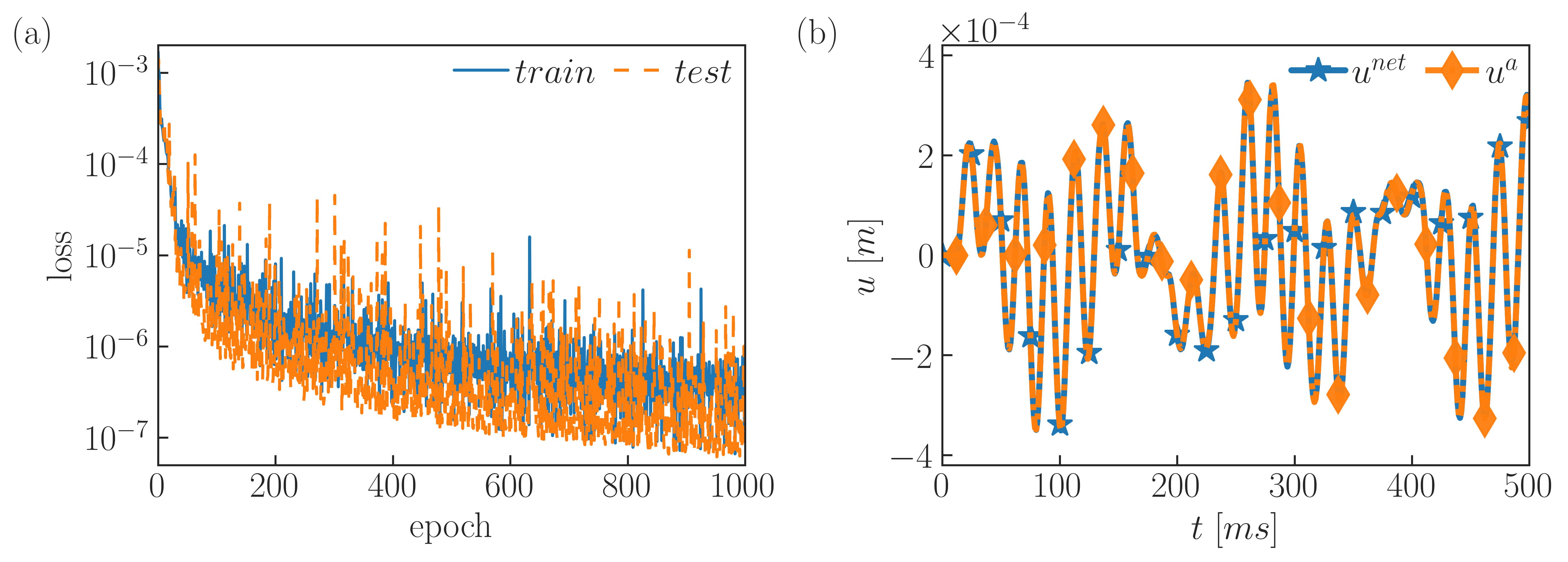}
	\caption{(a) Training and test loss trend versus training epochs for the network element representing the transient response of rods subject to boundary load and zero initial conditions. (b) Prediction of the network element for the sample problem $rod_1$. Results correspond to the response at $x=0.6~m$ for the excitation frequency $\omega_0 = 294.7~rad/s$, randomly selected from the test dataset. Superscripts $\Box^a$ and $\Box^{net}$ refer to the solutions obtained from the analytical solution and the network element predictions, respectively.}\label{fig: BV_rod_training_and_sample}
\end{figure}

Figure~\ref{fig: BV_rod_training_and_sample}a shows the network element training and test losses versus the training epoch.
Figure~\ref{fig: BV_rod_training_and_sample}b shows the response for a sample test case labeled $rod_1$. This sample case was randomly drawn from the test dataset, so its properties follow the descriptions provided above. Comparing the solution predicted by the network with the analytical solution of the problem shows that the network predicted the response of $rod_1$ with minimal error.

We further evaluate the network element prediction accuracy using the relative absolute percentage error metric (in this paper it is also referred to as \textit{relative error}) defined as:

\begin{equation}\label{eq: network metrics}
\begin{gathered}
		e_{r} = \frac{1}{n_t. n_x} \sum\limits_{i=1}^{n_t}{\sum\limits_{j=1}^{n_x}\frac{ \left| u^{net}(x_j,t_i) - u^\text{true}(x_j,t_i)\right|}{
    	\max\limits_{1<i<n_t, 1<j<n_x}\left\{ u^\text{true}(x_j,t_i) \right\} - \min\limits_{1<i<n_t, 1<j<n_x}\left\{ u^\text{true}(x_j,t_i) \right\} }}\times 100\% 
	%
\end{gathered}
\end{equation}

For the developed network element, the average train and test $e_r$ are $0.246~\%$ and $0.248~\%$, respectively.
Also, for the test sample $rod_1$, the relative error is $0.307~\%$.
The small $e_r$ value indicates the excellent agreement between the predictions of the network elements and the true (analytical) solution of the problem. We highlight that the test dataset was never used for training the network element and therefore was never seen by the network. Hence, the low percentage error shows the network is generalized enough to accurately predict the response of a class of boundary value problems subject to harmonic boundary load. 

\subsection{Case~2: Transient response to arbitrary initial conditions}\label{sec: initial BV rod}
As discussed in \S~\ref{sec: SFNE description}, the SFNE is equipped with two types of input channels ($In^D$ and $In^S$) to receive and process external inputs, properties, and initial and boundary conditions. In the previous section, we showed how a sample dynamic (time-evolving) input is given to the SFNE. The goal of this section is to show how nonzero initial conditions can also be provided as input to FENA.

\subsubsection{Problem statement:}\label{sec: probelm statement initial BV rod} 
Consider the problem presented in \S~\ref{sec: BV Rod Problem Statement}, which is a uniform rod subject to an axial boundary load as shown in Figure~\ref{fig: rods_schematic}a. While in the previous section the rod was assumed initially at rest, here we consider nonzero initial conditions. Hence, the governing equation and boundary conditions are the same as Eq.~\ref{eq: BV_rod}, while the initial conditions $u_0(x)$ and $\dot{u}_0(x)$ are nonzero. The goal is to find the deformation and velocity distribution of the rod for a given set of initial conditions and boundary load. 




\subsubsection{SFNE architecture}\label{sec: IV_rod_net_arch}

\noindent \textbf{Informing SFNE of initial conditions:} As discussed in \S~\ref{sec: SFNE description}, the initial conditions should be fed to the network through $In^S$, since the initial conditions are constant problem conditions and do not evolve with time. 
Hence, the initial states of the recurrent units of the BRNN block are adjusted according to the initial conditions. 
The recurrent units (the forward and backward LSTM units inside the BRNN) combine the initial states and $f(t_1)$ to predict the output at the first time step ($Out(x, t_1)$). Both the cell and the hidden states are updated, based on the initial states and $f(t_1)$, and will be used in the calculations at the following time step. Thus, the information used to initialize the BRNN block is combined with the dynamic input at each step and propagates across all time steps. 
This step is similar to the classical numerical methods, in which a computation is initialized based on a given set of initial conditions. While in this section, the only constant input is the initial conditions of the problem, in the following sections we will show how various types of constant parameters and conditions can be fed to the SFNE via $In^S$.

Similarly to the previous network element, for the initial value problem, we use separate MLP blocks ($\text{NN}_{FCS}, \text{NN}_{FHS}, \text{NN}_{BCS}$, and $\text{NN}_{BHS}$) to initialize both the cell and hidden states of the LSTM units independently of other states. It is worth highlighting that our experiments on possible architectures revealed that replacing $\text{NN}_{FCS}, \text{NN}_{FHS}, \text{NN}_{BCS}$, and $\text{NN}_{BHS}$ with a single MLP, which outputs a vector containing all the initial states of the forward and backward LSTM units, significantly reduces the accuracy of the network element. 
This is because using a shared MLP prevents the network from extracting various uncorrelated features needed to initialize each internal state of LSTM, hence preventing the SFNE from properly learning the dynamics of the problem. 

It is worth highlighting that as the width of a MLP network increases, the influence of each weight on the output can diminish, making the output scalars more uncorrelated. A single, larger MLP might be theoretically capable of learning to extract uncorrelated features (which in our case are the hidden states of the LSTM cells) given sufficient size and training. However, our numerical experiments suggest that in the practical application to FENA, using separate MLPs was more effective. More specifically, the use of separate MLPs ($\text{NN}_{FCS}, \text{NN}_{FHS}, \text{NN}_{BCS}$, and $\text{NN}_{BHS}$) resulted in superior performance, as observed through the comparison of network element accuracy. 
Although our observation is based on empirical studies, it highlights the practical advantage of using dedicated MLPs for initializing the LSTM units in the network elements.

\noindent\textbf{Network architecture:}
The internal architecture of the network element is the same as the network element of \S~\ref{sec: Case 1 SFNE Arch} (see Table~\ref{table: BV_rod_netArch}) except for the revised $In^S$ and $Out$ layers. \REV{$In^S$ is an array of size $\medmuskip=0mu 2\times n_x$, where the first row represents the initial deformation $u_0(x)$ on the grid of $n_x=51$ equally spaced nodes on the rod, and the second row is the initial velocity $\dot{u}_0(x)$ at the nodes.} The network element predicts both deformation and velocity (calculated at the aforementioned nodes) at each time step; hence the output size is $\medmuskip=0mu 2\times n_x$. Note that we extended the output to both deformation and velocity to allow concatenating multiple network elements in time and to predict the response beyond the maximum time steps, $t>t_{n_t= 100}$, which is the maximum time step that the network element sees during training (see also the discussions in \S~\ref{sec: IV_BV_rod time concat}).

\subsubsection{Network training and prediction results}\label{sec: initial value rod training}

For this problem, we considered the same boundary condition and material and geometry properties as in the previous section (\S~\ref{sec: Training_BV_Problem}). 
We used the data samples generated in \S~\ref{sec: Training_BV_Problem} to build the training data. Specifically, to obtain a training sample, we took a data sample from the dataset of \S~\ref{sec: Training_BV_Problem}, randomly selected an initial starting time $t_s\in[0,400]~ms$, and extracted the input force, displacement, and velocity for $t_s\leq t \leq t_{s+101}$ from the data sample. The displacement and velocity at $t_s$ provided the initial conditions, and the rest of the solution was used as the system response to the initial conditions and applied boundary load. The network element was trained to predict the response for the next hundred steps based on the boundary load and initial conditions.

Before we proceed further, we take the opportunity to clarify a few points about the proposed approach used to build the training dataset of this specific SFNE. The training dataset could be constructed using the available semi-analytical solution. This solution is based on series expansion and numerical integration of the initial conditions~\cite{doyle2020wave}, a solution approach that quickly becomes very computationally demanding.
However, the proposed approach to build the training dataset eliminates the computational cost because it is based on sampling from an available dataset. It follows that the required computational resources are very limited.
We also highlight that this dataset construction approach is inspired by data augmentation techniques in deep learning~\cite{bansal2022systematic_augmentation} and can be applied to any similar initial value problem where either the access to data is limited or generating training data is resource intensive.
Also, note that the selection of the \textit{prediction window} ($n_t=100$ time steps) of the network element was arbitrary. The 100 steps window was selected because it allowed selection of the initial time $t_s$ from a wide range of values within $[0,400]~ms$. On the other hand, a larger prediction window would limit our choice for the initial time step, as the data generated in \S~\ref{sec: Training_BV_Problem} is available only for up to 500 steps. 

We generated one sample per each data sample available in the dataset of \S~\ref{sec: Training_BV_Problem}. Therefore, the dataset size was $6\times10^4$. 
We used the same training hyperparameters and training algorithm as in the previous case (see \S~\ref{sec: Training_BV_Problem}). Note that this procedure can be repeated multiple times to generate different datasets needed to build an ensemble of SFNEs. In other words, for each data sample of \S~\ref{sec: Training_BV_Problem}, the initial time step can be chosen differently to obtain different nonzero initial condition data samples. 

\begin{figure}[!h]
	\centering
	\includegraphics[width=.99\linewidth]{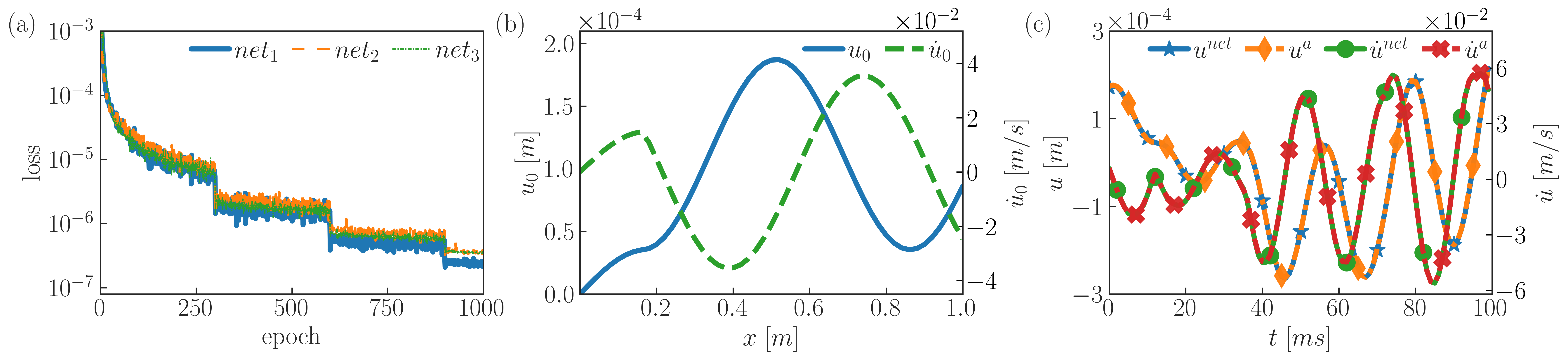}
	\caption{(a) Trend of network element ensemble training loss versus training epoch. $net_{1,2,3}$ refers to the SFNEs trained for the problem discussed in \S~\ref{sec: probelm statement initial BV rod}. (b) Initial conditions of $rod_2$. 
    (c) The response predicted by the network ensemble compared to the analytical solution of $rod_2$ for $x=0.6~m$. The response is presented in terms of displacement $u$ and velocity $\dot{u}$ profiles; the superscripts $\Box^{net}$ and $\Box^{a}$ indicate the solutions according to the SFNE and to the analytical approach, respectively.
    \label{fig: IV_rod_training_and_sample}
    }
\end{figure}

Using the procedure described above, three separate datasets were generated to train three SFNEs, and to form an ensemble of network elements.
Figure~\ref{fig: IV_rod_training_and_sample}a shows the trend of the training losses of the three SFNEs.
Note that the choice of the size of the ensemble was based on an experimental investigation in which we measured the gain in accuracy per added model to the ensemble. We noticed that while an ensemble of three network elements significantly reduced the average prediction error (compared to using a single network), a further increase in the size of the ensemble slightly improves the predictions, as seen in Figure~\ref{fig: rod_concat_sample}a. 
Note that while the trend of the loss function was monitored as a measure of overfitting, in the interest of brevity, we avoid presenting the test loss trend for both Case~2 and Case~3 and only report the relative error across the test dataset.

The average $e_r$ of the model ensemble calculated over the train and test data is $e_r = 0.194~\%$ and $e_r = 0.205~\%$, respectively.  
Note that the solution of each test problem is based on the average predictions of the ensemble of network element models. The small value of the relative error shows excellent agreement between the true solution available from the test dataset and the SFNE ensemble predictions, and it confirms the potential of the presented approach to solving initial value problems. 

To showcase the prediction performance of the network, we also present a sample problem solution obtained from the network element ensemble. We label this problem as $rod_2$, which is a rod with the properties defined in \S~\ref{sec: Training_BV_Problem}, subject to the harmonic boundary load with $\omega_0 = 294.7~rad/s$ (randomly selected from the test dataset), and the initial conditions shown in Figure~\ref{fig: IV_rod_training_and_sample}b. Figure~\ref{fig: IV_rod_training_and_sample}c compares the $rod_2$ response predicted by the network element ensemble with the analytical solution. For this problem, $e_r = 0.172~\%$. The results show that the network predictions are very close to the analytical solution of the problem, indicating the high prediction accuracy of the network element ensemble.

\subsubsection{Network concatenation: extending the SFNE prediction window}\label{sec: IV_BV_rod time concat}

In the previous section, we developed an ensemble of network elements that predicts the rod response up to $n_t = 100$ time steps for a set of initial conditions. In this section, the concatenation procedure discussed in \S~\ref{sec: concat description} is used to predict the response beyond the network prediction window of 100 steps. We follow the steps provided in Algorithm~\ref{Algorithm:concat} and evaluate the performance of the proposed method via the sample test problem presented in the following. 

\begin{figure}[!htbp]
	\centering
	\includegraphics[width=.8\linewidth]{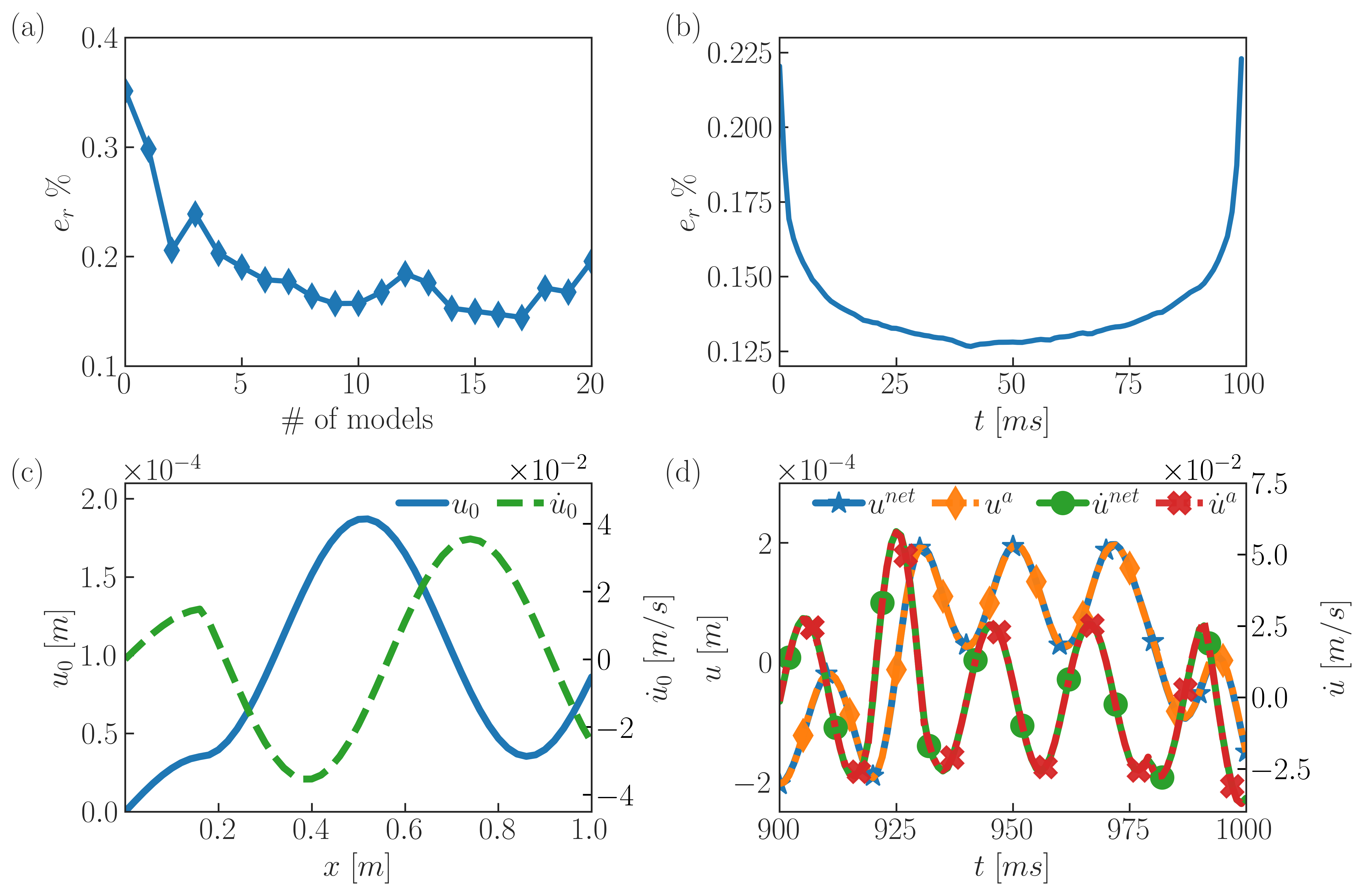}
	\caption{Network element concatenation.
	(a) Average model ensemble relative error versus the number of network element models in the ensembles. The test dataset of the first model in the ensemble was used to calculate the average $e_r$ of the ensemble. All 20 models in the ensemble have the same architecture but are trained with training datasets that are sampled differently. The same training hyperparameters are used for all the 20 network element models in the ensemble (see \S~\ref{sec: initial value rod training}).
	(b) Distribution of the network ensemble test $e_r$ versus the prediction time step. The relative error value remains low in the entire prediction time window. However, it gradually starts increasing when approaching the upper bound of the window. (c) Initial conditions of the sample case $rod_3$. (d) $rod_3$ response calculated by network element concatenation ($\Box^{net}$) compared with the response obtained from the analytical solution ($\Box^{a}$).
    }\label{fig: rod_concat_sample}
\end{figure}

The first step in implementing the concatenation approach is to develop an ensemble of SFNEs and to determine the size of the ensemble. We used the procedure described in \S~\ref{sec: initial value rod training}, to build and train a group of 20 SFNEs and form the average ensemble test $e_r$ against the number of models (SFNEs) within the ensemble (Figure~\ref{fig: rod_concat_sample}a). It is seen that with an ensemble of three network elements, there is a considerable decrease in $e_r$. It should be noted that although a larger number of models could further reduce the prediction error, the gain in accuracy is minor compared to the enhanced accuracy in an ensemble of three network elements. Hence, we used only the first three networks in the ensemble. 

The next step is to determine the cut-off threshold. This is done by forming the ensemble test $e_r$ versus the time-step graph shown in \S~\ref{fig: rod_concat_sample}b. It is seen that the error starts to increase rapidly at $75~ms$. Hence, the cut-off threshold step is set at $t_c=75$. 
Note that while the error is minimized around the $40~ms$, it remains in the same range until approximately $75~ms$.
It is worth highlighting that the higher error of the first and last few time steps ($t< 20~ms$ and $t>80~ms$ in Figure~\ref{fig: rod_concat_sample}b) can be attributed to a phenomenon commonly observed in sequential models like LSTMs. Considering the bidirectional flow of information within the SFNE, at the start of the sequence the forward LSTM has not yet seen any data; similarly, at the end of the sequence, the backward LSTM has not yet seen the data points from the proceeding time steps. Therefore, the predictions at these points rely only on the information from one direction. Consequently, these initial and final states lack contextual completeness compared to the middle states, which benefit from information gathered from both the preceding and succeeding time steps. This factor leads to relatively higher errors at the beginning and end of the sequence, resulting in the U-shaped error curve observed in Figure~\ref{fig: rod_concat_sample}b. 
It should also be noted that the rise seen in the $e_r$ trend is relative to the intermediate time steps ($t<75~ms$), and $e_r$ is small ($\max\{e_r\} < 0.23\%$) for the whole set of time steps.
After determining the size of the ensemble and $t_c$, the second part of the Algorithm~\ref{Algorithm:concat} should be followed to simulate the response beyond the prediction window of the SFNEs in the ensemble. 

To showcase the performance of the proposed concatenation algorithm, we applied it to the sample problem labeled as $rod_3$. The initial conditions for $rod_3$ are shown in Figure~\ref{fig: rod_concat_sample}c. The rod is subject to a harmonic boundary load with $\omega_0 = 294.7~rad/s$. The goal is to predict the rod response in the time window $0<t<1000~ms$.

To obtain the response for $0<t<1000~ms$, we first calculated the response up to $975~ms$ using 13 concatenation steps (i.e. $13 \times 75$~steps). We used the first 25 prediction steps of the last concatenation step (step 14, which provides the solution at $975<t<1075~ms$) and disregarded the rest of the predictions at the remaining time steps.
\REV{Figure~\ref{fig: rod_concat_sample}d shows the simulation results in the time range $900<t<1000~ms$ obtained via the concatenation process and compared to the analytical solution. It is seen that using the proposed concatenation method we accurately predicted the response up to $1000~ms$, that is, 10 times larger than the time window the network elements were trained to predict. The average prediction error for $rod_3$ is $e_r=0.203~\%$, indicating that the proposed algorithm enables extending the prediction window of the network elements.}

It is worth mentioning that the concatenation process cannot be performed indefinitely, as the accumulation of the prediction error eventually results in a deterioration of the prediction accuracy. However, with the proposed algorithm, the error propagation can be well controlled, and the concatenation algorithm predictions remain accurate for simulation periods much longer than the training prediction window of SFNEs. 
In practical applications, one way to handle this limitation of the concatenation process is to carefully design the SFNE model to reduce the number of concatenations based on the desired simulation period. Note that by selecting an appropriate prediction window for SFNE and limiting the number of concatenations, prediction accuracy can be maintained over the desired simulation period. An important factor in this process is the maximum window of the training data, which limits the prediction window of the SFNE. 
We also highlight that transient simulations are very computationally intensive. The ability to develop network models that can accurately predict the response much beyond the time range observed in the training dataset is quite notable and has many potential applications.
This has numerous potential applications, including the development of surrogate models based on early stages of simulations or experimental measurements. By using the available (simulation or experimental) data within a certain period of time, SFNEs can learn the system's dynamics and replicate its transient response. With the proposed algorithm, concatenation can extend the prediction horizon beyond the training data, thus reducing or eliminating the need for further resource-intensive simulations or measurements. Therefore, the proposed methodology can significantly reduce the computational/experimental burden and associated costs.

\subsection{Case~3: Transient simulation of rods subject to distributed load}\label{sec: dist load rod}

In this section, we develop network elements that represent a rod subject to a distributed load with nonzero initial conditions. Since the general procedure to develop the network elements is similar to the previous sections, in the following we omit the details and focus only on the main differences in the implementation of the SFNEs for Case~3.

\noindent\textbf{Problem statement:} Consider a thin rod aligned along $x$-axis with length $L$, uniform cross-section $A$, Young's modulus $E$, and density $\rho$. Both ends of the rod are connected to springs having stiffness $k_1$ and $k_2$), respectively; we will refer to this as spring boundary conditions as shown in Figure~\ref{fig: rods_schematic}b. The governing equations are: 

\begin{equation}\label{eq: spring_supported_rod}
\begin{gathered}
    E A \frac{\partial^2 u}{\partial x^2} +f(x,t) = \rho A \frac{\partial^2 u}{\partial t^2}\\
    \begin{aligned}
    u(x,0)&=u_0(x) &\dot{u}(x,0) &= \dot{u}_0(x)\\
    EA \frac{\partial u(0,t)}{\partial x} &= k_1 u(0,t) & EA \frac{\partial u(L,t)}{\partial x} &= k_2 u(L,t)
\end{aligned}
\end{gathered}
\end{equation}

\noindent 
We assumed that the rod is subject to a uniform distributed harmonic load $f(x,t)= f_0 sin(\omega_0 t)$; once again, this load was chosen because enabling a simple analytical solution. In addition, the case of spring boundary conditions was chosen to show that the proposed concept can learn different forms of boundary conditions, and it is not limited to the specific type of boundary conditions in \S~\ref{eq: BV_rod}. Similarly to the previous section, the goal is to develop network elements that serve as surrogate models capable of predicting the response over a uniform grid of $n_x$ points and $n_t$ time steps. In the interest of brevity, we will omit the zero initial condition case, whose implementation is similar to the nonzero initial condition case discussed in the following (the only difference being that $In^S= 0$ for zero initial conditions). 

\noindent \textbf{Training dataset:} 
We follow the same procedure as discussed in \S~\ref{sec: initial value rod training} to build the training dataset. First, a training dataset with zero initial conditions is generated. Then, the training data for the nonzero initial condition problem is built by taking samples from the zero initial condition dataset. 
For the spring boundary conditions, we used the same material and geometric properties as in \S~\ref{sec: Training_BV_Problem} (see Table~\ref{table: rod Params}). We also assumed $f_0=1$, $k_1 = EA/2L$, and $k_2 = 2EA/L$. Note that the boundary stiffness values were arbitrarily chosen, and any other combination of boundary stiffness values does not limit the application of FENA to the problem. A total of $6\times 10^4$ samples were generated by randomly selecting $\omega_0$ from the range $[52.5, 366.5]~rad/s$ and with zero initial conditions. Each data sample includes the displacement and velocity of the rod for $t\in [0,500]~ms$ with a time step of $1~ms$. Data samples were generated using the analytical solution of the system described below: 

\begin{equation}\label{eq: sol_dist_load_rod}
\begin{gathered}
u(x,t) = \sum\limits_{r=1}^{\infty} 
\frac{f_0 B_i U_i(x)}{\rho A \omega_i}
\frac{\omega_0 \sin(\omega_r t) - \omega_r \sin (\omega_0 t)}{\omega_0^2-\omega_r^2}\\
U_i(x) = \cos \left(\frac{\omega_i x}{l}\right) + \frac{k_1 c}{E A \omega_i} \sin \left(\frac{\omega_i x}{l}\right), B_i = \frac{\int_0^l{U_i(x)dx}}{\sqrt{\int_0^l U_i^2(x)dx}} 
\end{gathered}
\end{equation}

\noindent where $\omega_i$ is the $i^{th}$ natural frequency of the rod. Further details of the analytical solution can be found in~\cite{doyle2020wave}. To build the training dataset for the network elements, we extracted data samples from the zero initial condition data, following the procedure discussed in \S~\ref{sec: initial value rod training}.

\noindent \textbf{SFNE architecture:} For the spring boundary conditions, we used again the network element architecture described in \S~\ref{sec: IV_rod_net_arch}. Recall that the external applied load is uniform, hence it is represented by a single value at each time step ($In^D(t_i) = f_0 \sin (\omega_0 t_i)$); the network elements learn that $In^D$ that represents the amplitude of the uniform load. Note that the SFNE is not limited to uniform external loads. In the case of a spatially variable load, the load can be mapped onto a vector of nodal loads using the concepts of FEA~\cite{reddy2006theory}. 
Specifically, at each time step, FEA is used to calculate the equivalent resulting nodal loads at the discrete spatial locations $\{x_1, ..., x_{n_x}\}$. The resulting nodal loads form a vector that is provided to the network as $In^D$. 

\begin{figure}[!h]
	\centering
	\includegraphics[width=.99\linewidth]{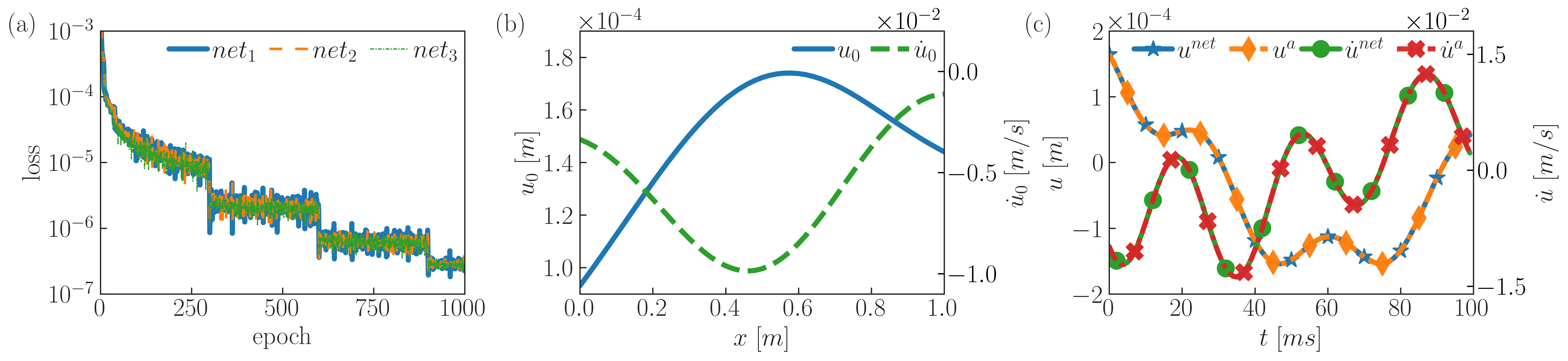}
	\caption{(a) The trend of training loss of the network elements developed for the simulation of rods subject to distributed load and nonzero initial conditions. (b) Initial conditions of the sample $rod_4$. (c) The average response of $rod_4$ predicted by the ensemble compared to the analytical solution. The results correspond to the physical location at $x=0.6~m$. 
 }\label{fig: Dist_rod_training_and_sample}
\end{figure}

\noindent \textbf{Prediction results:} An ensemble of three network elements was trained using the same training hyperparameters as in \S~\ref{sec: Training_BV_Problem}. The trend of training loss is shown in Figure~\ref{fig: Dist_rod_training_and_sample}a.
The average relative train and test error of the ensemble are $0.165~\%$ and $0.176~\%$, respectively, showing that the ensemble has a minimal prediction error across the training and test datasets.

We also present a sample prediction to display the performance of the developed network element ensemble. The sample problem is labeled as $rod_4$. The rod is subject to a uniform distributed harmonic load with $\omega_0= 185.48~rad/s$ and the initial conditions shown in Figure~\ref{fig: Dist_rod_training_and_sample}b.
Figure~\ref{fig: Dist_rod_training_and_sample}c shows the deformation of $rod_4$ calculated using the network elements and the analytical solution (Eq.~\ref{eq: sol_dist_load_rod}). The prediction error for $rod_4$ is $0.114~\%$. The results show that the network elements accurately predicted the transient response of the rod. Note again that the reported $e_r$ and $rod_4$ results are based on the average prediction of the three network elements in the ensemble.

\subsubsection{On-demand prediction location}\label{sec: on-demand location}

In the previous set of results, all the network elements predicted the displacement and velocity distributions at a fixed set of points. However, one might require to obtain the output at a specific spatial location that could fall between two consecutive nodes. One option to address this problem is to use standard interpolation techniques following classical numerical methods, such as FEA~\cite{reddy2006theory}. However, the proposed SFNE architecture offers the opportunity to develop networks that predict the response at an arbitrary location $x$ instead of a fixed grid of points. The development of these networks follows very similar steps to those taken to develop a transient network element. Specifically, to build a network element that predicts the response of the rod at an arbitrary point $x=x_0 \in [0, L]$ for a given set of initial conditions and applied distributed load, we follow the following steps:
\begin{itemize}
    \item \textit{SFNE architecture:} The internal network architecture used in the above section can also be used in this case. However, the input and output layers should be revised. $In^D$ is still defined as the time history of the applied load, hence $In^D_i = f(t_i)$. The output vector will be revised to a vector of size 2 defined as $Out_i(t_i) = \{u(x_0, t_i), \dot{u}(x_0, t_i)\}$. Note that the location at which we seek the response ($x_0$) is the same for all time steps. Thus, it is categorized as a constant input. As discussed previously, constant inputs should be provided to the network through $In^S$. Hence, we will add an extra channel to the $In^S$ layer and feed the value of $x_0$ to the network element. In other words, in this case, $In^S$ is composed of three parts: 1) $u_0(x)$ 2) $\dot{u}_0(x)$, and 3) $x_0$. 
    \item \textit{Training datasets:} The training dataset for this case can also be sampled from the data currently available, that is the dataset generated for Case~3. Specifically, we break every data sample into $n_x$ data samples, which are the time history of the response at the $n_x$ number of nodes. The dataset is then randomly split to train and test datasets with a split ratio of 85\% / 15\%.
    \item \textit{SFNE training:} The networks were trained using the same training scheme discussed in \S~\ref{sec: Training_BV_Problem}. 
    
\end{itemize}

\begin{figure}[!h]
	\centering
	\includegraphics[width=\linewidth]{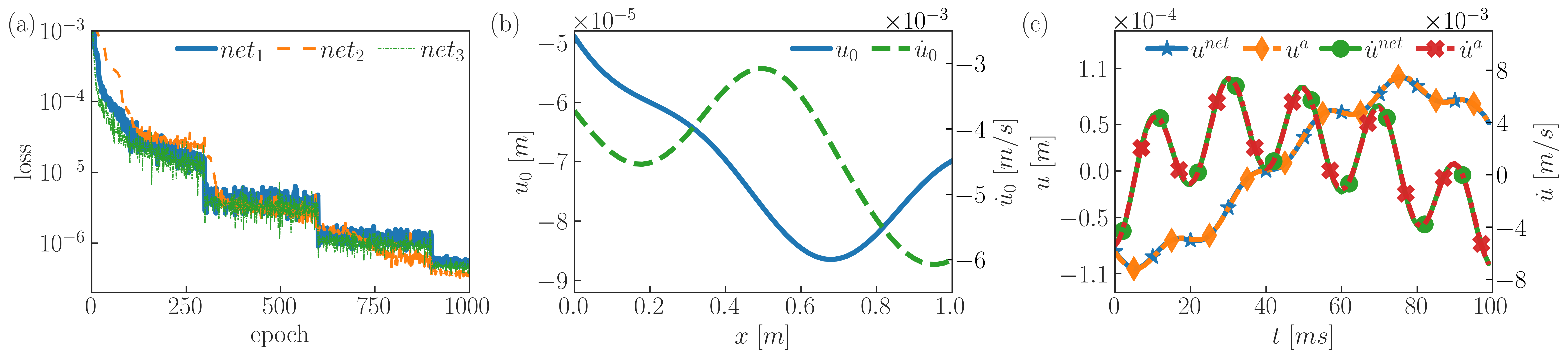}
	\caption{(a) Training loss of network elements with auxiliary $x$ input channel. (b) Initial conditions of the sample problem $rod_5$. (c) Displacement and velocity of $rod_5$. The rod is subject to a harmonic distributed load with $\omega_0 =308.1~rad/s$. The results correspond to the axial location $x=0.81~m$.
    }\label{fig: X_input_rod_training_and_sample}
\end{figure}

Figure~\ref{fig: X_input_rod_training_and_sample}a shows the trend of loss function of the network elements with the $x_0$ input channel during training. Again, we trained an ensemble of three network elements. The average prediction error across the training dataset is $0.355~\%$. Also, the average prediction error of the test samples is $0.389~\%$, showing that the network has learned to accurately predict the response at an arbitrary location. %

We also show a sample prediction of the network element ensemble in Figure~\ref{fig: X_input_rod_training_and_sample}c. The response corresponds to a sample problem labeled $rod_5$. The rod governing equations are given by Eq.~\ref{eq: spring_supported_rod}, and its material properties are the same as in \S~\ref{sec: Training_BV_Problem}. The initial conditions of $rod_5$ are shown in Figure~\ref{fig: X_input_rod_training_and_sample}b. The goal is to find the response at $x = 0.81~m$, a (randomly selected) location that is not a member of the training dataset. 
The relative prediction error for $rod_5$ is $0.441~\%$ (calculated with respect to the analytical solution). Results show that the SFNE ensemble accurately predicts the response at the desired location, although the network elements had never seen $x=0.81m$ during training. The accurate predictions also show that the network elements have learned the dynamics of the problem and are able to correctly simulate the response at any desired location. 

The case presented here further demonstrates the generalizability and flexibility of the concept of SFNE. It also describes, through an example problem, how additional parameters can be provided to network elements. In the following sections, we will discuss how this approach can be followed to inform network elements about inhomogeneities of the computational domain. 

\subsection{Remarks on the computational cost and benefits of FENA}\label{sec: computational costs}

In the following, we discuss the computational aspects of FENA and compare them with the cost associated with other solution approaches. 
Generally, there are two main computational aspects associated with the application of FENA to the transient simulation of physical systems: 1) the cost for the development of the network elements, and 2) the cost of simulations with the trained network elements. We discuss the costs associated with each aspect here below.
\begin{enumerate}
    \item The cost for the development of network elements consists of two main parts, the cost of data generation and the cost of the network element training. The computational cost associated with this step appears to be high because obtaining training data samples and training network elements are typically computationally expensive steps. Nevertheless, we should consider that, in FENA, this step is performed only once. Once the library of SFNE is available, it can be used for predictions with very minimal computational effort. Hence, the computational effort associated with the development of SFNE becomes of marginal importance in the whole process. 
    The training data for the above-discussed problems were generated on a cluster node with an NVIDIA Tesla V100 GPU, a Xeon Gold 5218R CPU with 40 cores at 2.10GHz base frequency, and 192 GB of memory. The average computational time required to generate a data sample with zero initial conditions is $0.11~s$. We generated each training dataset in less than 60 minutes applying parallel computation techniques and utilizing all 40 available cores. We also highlight that data generation for the cases with nonzero initial conditions is not computationally expensive since such datasets are built simply by taking samples from the readily available data samples (with zero initial conditions). Note also that the data generation in the above cases is not very computationally expensive because we use analytical solutions to build training and test data. Of course, when analytical solutions are not used, the computational time will tend to rise, but this will not change the fact that the process will only be performed once.
    In addition, since all network elements have similar structures and training data formulation, they all have similar training computational costs. Using the cluster node GPU, the SFNEs with arbitrary initial conditions, on average, required 5.5~$hr$ of training time, and the training time for the SFNE of Case~1 (zero initial conditions) was 13.8~$hr$. The training time of SFNE of Case~1 is greater because it is trained over longer sequences and requires more time for backpropagation and weight updates in each training iteration.

    \item In the prediction step, the SFNE of Case~1 took an average of $0.0012~s$ to predict a response. Also, on average, the networks with arbitrary initial conditions predict a response in $0.0002~s$. These numbers are calculated on the basis of the average time of predictions for the test samples. On the other hand, the analytical solutions take an average of $0.11~s$ to provide the solution. The numbers indicate that the network elements reduce the computational time by at least two orders of magnitude when compared to the analytical solution. Note that even though the analytical solution is available for all the case studies in this section, they are series-based solutions and require a large number of numerical operations and calculations. It is clear that even this class of analytical solutions may quickly become computationally intensive, which justifies the significant computational speed-up seen when network elements are used to calculate the response in all the cases discussed in this section. Further, while the simulation process in numerical solutions (e.g. finite differences, finite elements) rapidly becomes computationally expensive for complex problems (not admitting analytical solutions), for neural networks the computational time does not necessarily scale up with the complexity of the problem. This computational benefit will be evident in the following example of simulation of inhomogeneous beams. 
\end{enumerate}

\section{Transient simulation of inhomogeneous physical systems}\label{sec: beam}
To further demonstrate the flexibility of the proposed approach, we apply FENA and the concept of SFNE to the transient simulation of inhomogeneous computational domains. Specifically, we focus on the transient simulation of beams governed by a higher-order PDE and under the assumption of nonuniform material and geometric properties. We discuss how the input $In^S$ is leveraged to inform the network element about the material and geometric properties. 

\subsection{Case~4: Inhomogeneous beams with zero initial conditions}\label{sec: Zero_init_beam}
\subsubsection{Problem statement}
\begin{figure}[!h]
	\centering
	\includegraphics[width=0.8\linewidth]{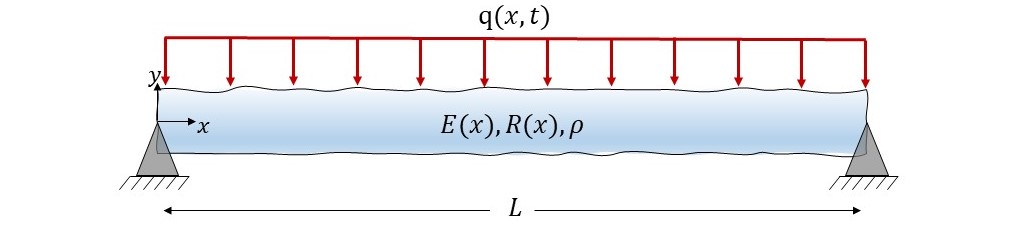}
	\caption{\REV{Schematic illustration of the proposed inhomogeneous beam for \textbf{Case-4 and Case-5} with spatially varying Young's modulus $E(x)$, non-uniform circular cross-section with radius $R(x)$, and uniform density $\rho$. The beam is subject to uniformly distributed harmonic transverse load $q(x,t)$. Note that cases-4 and 5 differ in their consideration of initial conditions.}}\label{fig: beams_schematic}
\end{figure}

Consider a slender beam of length $L$ where its neutral axis is aligned along the $x$-axis and has pin supports at $x=0$ and $x=L$. We assumed that the beam has a nonuniform circular cross-section with radius $R(x)$, spatially varying Young's modulus $E(x)$, and uniform density $\rho$ as shown in Figure~\ref{fig: beams_schematic}. The beam is subject to a uniformly distributed harmonic transverse load $q(x,t) = q_0 sin(\omega_0 t)$ assumed to act along the $y$-axis. The goal is to calculate the transverse deformation of the beam in the $y$ direction ($v(x,t)$) resulting from the applied transverse load. The beam governing equation is as follows:

\begin{equation}\label{eq: beam_governing_eq}
\begin{gathered}
    \frac{\partial^2}{\partial x^2} \left( E(x)I(x) \frac{\partial^2 v}{\partial x^2}\right) + \rho A(x) \frac{\partial^2 v}{\partial t^2} = q(x,t)\\
    \begin{aligned}
    & v(x,0)=v_0(x) &  \dot{v}(x,0)= \dot{v}_0(x) & \\
        v(0,t)=0 &\qquad v(L,t)= 0 &\quad \frac{\partial^2 v(0,t)}{\partial x^2}= 0  \qquad& \frac{\partial^2 v(L,t)}{\partial x^2}=0\\ 
    \end{aligned}
\end{gathered}
\end{equation}

In the following, we first focus on developing network element models for the beam simulation with zero initial conditions ($v_0(x) = \dot{v}_0(x)=0$). Then, in \S~\ref{sec: IV beam} we develop the network elements for arbitrary nonzero initial conditions. We highlight that, although the following discussions are cast in the form of an example problem (i.e. the inhomogeneous beam), the instructions below are general and applicable to any similar transient problem. 

\subsubsection{SFNE architecture}\label{sec: beam SFNE architecture design}
Following a procedure similar to that used in the previous section to design the beam network element:
\begin{itemize}
    \item \textit{$In^D$ and $Out$:} similarly to \S~\ref{sec: dist load rod}, $In^D$ at the $i^{th}$ time step is defined as $In^D_i = q|_{t=t_i}$, that is the amplitude of the uniformly distributed harmonic load at $t=t_i$. Therefore, the size of the $In^D$ layer is one.
    The network element is set to predict the beam transverse deformation and velocity, that is $v(x,t)$ and $\dot{v}(x,t)$ on $n_x$ equally spaced nodes at $t_1, t_2, ..., t_{n_t}$. Hence, the output layer size is $\medmuskip=0mu 2\times n_x$. We also highlight that any other parameter associated with deformation (such as stress components or beam rotation) is considered a dependent parameter and can be calculated via numerical derivation~\cite{reddy2005FEM} or neural network-based post-processing computations~\cite{jokar2022two}. 
    \item \textit{$In^S$:} The beam cross-section radius and Young's modulus are properties that depend only on location $x$ and do not vary over time, so they are classified as constant inputs. Hence, $R$ and $E$ should be defined and passed to the network through $In^S$. Specifically, we provide the distribution of these two properties as a vector representing the values of $R(x)$ and $E(x)$ on the grid of nodes at which network outputs are predicted. Hence, the size of the vector $In^S=\{R(x), E(x)\}$ is $\medmuskip=0mu 2\times n_x$. Note again that we do not need to feed the zero initial conditions to the network element 
    because, during training, the network implicitly learns that the beam starts from rest. 
    \item \textit{$\text{NN}_{FS}$ and $\text{NN}_{BS}$:} Given that $In^S$ is defined as a vector representing the distribution of $E$ and $R$, we can still use MLP blocks in $\text{NN}_{FS}$ and $\text{NN}_{BS}$ to preprocess $In^S$. However, this approach faces a limitation where a very fine grid of points (large $n_x$ values) is required to describe the distribution of properties in the beam. Note that the size of the MLP input layer must be equal to the length of the vectors $E$ and $R$. Thus, the input layer size scales up with the resolution needed to describe the nonuniform properties of the beam. Here, we adopt a different approach to address the aforementioned shortcoming by first processing $In^S$ via convolutional neural networks (CNN). Recall that the CNN architecture is based on convolutional kernels (filters) that slide along the input and scan different input regions. The kernels extract the local features in different regions of the input and map them to an array called feature map. Since CNNs use a set of fixed-sized convolutional filters to scan different regions of the input, the size of the CNN filters is independent of the resolution (size) of its input channel~\cite{Goodfellow-et-al-2016}. We attached the last layer of CNN to a series of fully connected layers (MLP) to further process the extracted feature maps before feeding them as initial states to the BRNN.
    \item \textit{BRNN:} For the BRNN block, again we use LSTM cells as forward and backward recurrent cells. 
    \item \textit{$\text{NN}_{{out}}$:} We use a MLP to further process the output of the BRNN block. However, we used a revised activation function to compensate for the increased intricacy of the inhomogeneous beam behavior. As discussed above, one of the main challenges in neural network-based transient simulations is that neural networks are prone to fail to learn rapid transitions in the output channel. This is due to the fact that neural networks tend to converge to the low-frequency components (or moving average) of the output rather than the fine details originating from the high-frequency components~\cite{wang2021understanding, ziyin2020neural}; this shortcoming may prevent training convergence or produce overfitting on the training dataset. Note that using $\sin(x)$ (or $\cos(x)$) as the activation function is not a solution, since using harmonic functions results in many local minima in the loss function and, therefore, could impede the convergence of training (optimization)~\cite{ziyin2020neural}.
    In order to allow the network to learn high-frequency components of the output signal, for all layers of $\text{NN}_{out}$ we use the Snake activation function proposed by Ziyin \etal~\cite{ziyin2020neural} defined as follows: 
    \begin{equation}\label{eq: revised-activation}
	Sn(x) = x + \frac{\sin(a_1x)}{a_1} 
    \end{equation}
    where $a_1$ is a trainable parameter learned during training. Our numerical observations indicated that the Snake activation function results in improved training convergence, reduced overfitting, and the highest prediction accuracy.
\end{itemize}

Table~\ref{table: BV_rod_netArch} summarizes the beam network element architecture. The core BRNN block includes 100 LSTM cells in each of the forward and backward directions. The material and cross-section properties ($E$ and $R$) are first processed via convolutional neural networks, labeled as $\text{CNN}_{S_{1-4}}$. $\text{CNN}_{S_{1-4}}$ output is passed to the MLP blocks $\text{NN}_{FHS}$, $\text{NN}_{BHS}$, $\text{NN}_{FCS}$, and $\text{NN}_{BCS}$, respectively. 

\begin{table}[!h]
	\caption{Summary of network element architecture for the simulation of the inhomogeneous beam. The activation function $Sn$ is defined in Eq.~\ref{eq: revised-activation}. $\text{NN}_{FHS}$, $\text{NN}_{BHS}$, $\text{NN}_{FCS}$, and $\text{NN}_{BCS}$ are composed of a CNN network (labeled as $\text{CNN}_{S_{1-4}}$) followed by a MLP block. The size and AF reported for $\text{NN}_{FHS}$, $\text{NN}_{BHS}$, $\text{NN}_{FCS}$, and $\text{NN}_{BCS}$ correspond to the MLP block. The architecture of $\text{CNN}_S$ is presented in Table~\ref{tab: CNN1_net_architec}.
	\label{table: Beam_netArch}}
	\centering
	\resizebox{.7\columnwidth}{!}{%
    \begin{tabular}{|c|c|c|c|c|c|}
        \hline
         \multirow{3}{*}{$\text{NN}_{FHS}$} & type&  $\text{CNN}_{S_1}$+MLP & \multirow{3}{*}{$\text{NN}_{BHS}$}& type& $\text{CNN}_{S_2}$+MLP \\
         & size& $\{100,100\}$ & & size & $\{100,100\}$\\
         & AF  & \{$tanh, ES$\} & & AF & \{$tanh, ES$\}\\
         \hline
        \multirow{3}{*}{$\text{NN}_{FCS}$} & type&  $\text{CNN}_{S_3}$+MLP &  \multirow{3}{*}{$\text{NN}_{BCS}$}& type& $\text{CNN}_{S_4}$+MLP \\
         & size& $\{100,100\}$ & & size & $\{100,100\}$\\
         & AF  & \{$tanh, ES$\} & & AF & \{$tanh, ES$\}\\
         \hline
        \multirow{3}{*}{$\text{NN}_{In^D}$} & type& MLP & \multirow{3}{*}{$\text{NN}_{{out}}$}& type& MLP\\
         & size& $\{50,150,150,150\}$& &size & $\{200,200,200,200,200,200,200\}$\\
         & AF& \{$tanh, ES, ES, ES$\} & & AF & \{$Sn, Sn, Sn, Sn, Sn, Sn, Sn$\}\\
         \hline
    \end{tabular}
	}
\end{table}

\begin{table}[!h]
\caption{\label{tab: CNN1_net_architec} Details of $\text{CNN}_{S_{1-4}}$ architecture. In the first three convolution layers (layers \#1, 4, and 7), the stride is 1. In the last convolution layer (layer \#10) and the max-pooling layers, the stride is 2. In dropout layers, the kernel size reports the dropout rate. (Conv.: 1D convolution, Pool.: max pooling, Drop.: dropout)}
\centering
\resizebox{.7\columnwidth}{!}{%
\begin{tabular}{|c|c|c|c|c|c|c|c|}
\hline
layer $\#$ & layer& kernel size & $\#$ of kernels &layer $\#$ & layer & kernel size & $\#$  of kernels \\ \hline
1         & Conv. & $3\times1$  & 8             &7          & Conv. & $5\times1$  & 16  \\ \hline
2         & Pool. & $2\times1$  & -             &8          & Pool. & $2\times1$  & -   \\ \hline
3         & Drop. & $.2$        & -             &9          & Drop. & $.2$        & -   \\ \hline
4         & Conv. & $3\times1$  & 16            &10         & Conv. & $5\times1$  & 16  \\ \hline
5         & Pool. & $2\times1$  & -             &11         & Pool. & $2\times1$  & -   \\ \hline
6         & Drop. & $.2$        & -             &12         & Drop. & $.2$        & -   \\ \hline
\end{tabular}
}
\end{table}
\subsubsection{Training data}
Given that an analytical solution is not available for this problem, we used the commercial finite element (FE) software COMSOL Multiphysics to generate training data samples. The model was built via the beam element available in the COMSOL structural module. The material and geometric properties used for the beam are summarized in Table~\ref{table: beam Params}. Note that the choice of functions $E(x)$ and $R(x)$ was arbitrary, and any other type of function could be used. 
In fact, any spatially varying parameter can be given to the network in the form of a vector that describes the spatial distribution of the parameter within the beam. For example, consider the case of a porous beam with random porosity. Although the distribution of pores within the beam is random, it is possible to model the inhomogeneity of the beam with a vector representative of the spatially varying porosity~\cite{patnaik2022role}. The resulting vector can be used as $In^S$ of the SFNE and to develop a surrogate network element model of the porous beam. It is also worth mentioning that $In^S$ could also be defined based on $\{\omega_r, \omega_E\}$ since the network element can learn the resulting pattern of radii and Young's moduli and their effect on the beam deformation distribution. However, defining $In^S$ based on a vector of the spatial distribution of $R$ and $E$ is more general and can be used for any beam problem with unknown $E(x)$ and $R(x)$ functions. 

Further, a small amount of damping was added to the system in order to improve the numerical stability of the FE model. Note that the inclusion of damping avoids the unrealistic unbounded response when the forcing frequency matches the resonance frequency, and accounts for a more realistic system behavior, given that all physical systems have some amount of damping. The damping was formulated in Rayleigh form given in the following: 
\begin{equation}\label{eq: }
f_d =\alpha_d \rho A \frac{\partial v}{ \partial t} + \frac{\partial}{\partial x}\left(\beta_d \frac{\partial}{\partial t} \left( \frac{\partial M}{\partial x} \right) \right) 
\end{equation} 
\noindent where $M$ is the internal bending moment, $f_d$ is the transverse damping load, $A$ is the beam cross-section area, and the mass and stiffness parameters are $\alpha_d=1.13~1/s$ and $\medmuskip=0mu \beta_d =1.31\times10^{-5}~s$, respectively.

\REV{Each data sample is defined by randomly selecting the values for $\omega_E$, $\omega_r$, and $\omega_0$. We sampled the combination of $\{\omega_E, \omega_r, \omega_0\}$ from the specified ranges via Latin Hypercube Sampling (LHS) method to ensure that all areas of the parameter space were appropriately sampled~\cite{mckay2000comparison}}. We generated a total of $10^4$ samples and simulated each sample via COMSOL. Each solution was calculated for $n_x = 101$ evenly spaced nodes over a $0<t<400~ms$ time window and with a time step of $1~ms$. The number of nodes was increased in order to avoid aliasing effects. This was necessary because due to the higher number of degrees of freedom and the presence of inhomogeneities the solution of the beam equation exhibits relatively higher frequency oscillations, compared to the rod problem in \S~\ref{sec: rod}, and requires a higher number of nodes for an accurate representation of the beam response. 
The number of training samples was reduced compared to the cases discussed in \S~\ref{sec: rod} in order to show that the SFNE can still learn a relatively more intricate pattern despite a reduction in the number of training samples. Additionally, the maximum time step was decreased (compared to the case of the rod with zero initial conditions) in order to reduce the computational cost of data generation.

\begin{table}[!h]
 \caption{Dimensions and material properties used for the beam example.\label{table: beam Params}}
  \centering
  \resizebox{.9\columnwidth}{!}{%
  \begin{tabular}{|c|c|c|c|c|c|}
    \hline
    $L$ & $\nu$ & $\rho$ & $R(x)$& $E(x)$ & $q(x,t)$ \\
    \hline
    \multirow{3}{*}{$0.5~m$} & \multirow{3}{*}{$0.3$}& \multirow{3}{*}{$2000~kg/m^3$}& $R_0\left(1+ \sin^4(\omega_r x)\right)$ & $E_0 \left(1+.3\cos^4(\omega_E x) \right)$& $q_0 \sin(\omega_0 t)$\\
     & & & $R_0 = 0.01~m$&  $E_0 = 10^7~Pa$& $q_0 = 1~N$\\
     & & & $\omega_r \in \left[\frac{\pi}{L}, \frac{10 \pi}{L}\right] $ & $\omega_E \in \left[\frac{\pi}{L}, \frac{10 \pi}{L}\right] $ & $\omega_0 \in [6.28, 622.04]~rad/s$ \\
    \hline
  \end{tabular}
  }
\end{table}

\subsubsection{Network training and prediction results}\label{sec: training_zero_init_beam_network}
For the beam network elements, we revised the initial learning rate, type of learning rate scheduler, loss function, and the total number of training epochs compared to the training scheme described in \S~\ref{sec: Training_BV_Problem}, as described in the following:
\begin{itemize}
    \item \textit{Learning rate:} The initial learning rate was set to $0.0002$ as we observed that higher learning rates resulted in divergence in network training. 
    \item \textit{Training epoch:} We increased the total number of training epochs to 1500 to compensate for the slower training speed caused by the reduced learning rate. 
    \item \textit{Learning rate scheduler:} For the rod network elements, we used a scheduler that reduces the learning rate by a factor of two every 300 epochs. We observed that this scheduler might result in overfitting on the training dataset because the learning rate might become too small and the optimizer converges to a local minimum. 
    Hence, we used another scheduler that reduces the learning rate only if the loss does not improve further. Specifically, the scheduler only reduces the learning rate by a factor of two if the loss does not improve for 75 epochs. A shorter patience period resulted in reduced learning rates in the early stages of training and prevented convergence.
    \item \textit{Training loss:} We observed that the MSE loss tends to converge to the moving average of the output rather than fully capturing the output response. This is mainly because of the different scales of training samples, which cause the training algorithm to overlook (relatively) minor absolute errors in some of the samples while the relative prediction error is still large (in those samples). To overcome this issue, we revised the loss function to: 
    \begin{equation}\label{eq: revised loss}
	\mathcal{L} = \frac{\mathcal{W}_1}{N} \sum\limits_{i=1}^{N}\left[\left(v_i^{net} - v_i^{FEA}\right)^2 + \left(\dot{v}_i^{net} - \dot{v}_i^{FEA}\right)^2\right] + \frac{\mathcal{W}_2}{N} \sum\limits_{i=1}^{N}\left[\frac{\left(v_i^{net} - v_i^{FEA}\right)^2}{rng(v_i^{FEA})^2} + \frac{\left(\dot{v}_i^{net} - \dot{v}_i^{FEA}\right)^2}{rng(\dot{v})^2}\right]
    \end{equation}    
    where $\mathcal{W}_1$ and $\mathcal{W}_2$ are the weight factors of the loss function terms, $v_i=v_i(x,t)$ is the spatiotemporal deformation of the beam in the data sample $i$. The superscripts $\Box^{net}$ and $\Box^{FEA}$ refer to the SFNE prediction and FEA solution (available in the training dataset), respectively. The function $rng(.)$ calculates the range of each data sample and is defined as $rng(v_i) = max\{v_i(x,t)\} - min\{v_i(x,t)\}$. The weight factors $\mathcal{W}_1$ and $\mathcal{W}_2$ are initialized to 1 and optimized during training. Specifically, the training process can be expressed as:
    \begin{equation}\label{eq: revised training}
    \max_{\mathcal{W}_1, \mathcal{W}_2}\min_{\mathbf{\Theta}}~ \mathcal{L}\left(\mathcal{N}(In^D, In^S;\mathbf{\Theta}), {Out}^{true}(x,t); \mathcal{W}_1, \mathcal{W}_2\right)
    \end{equation} 
    During training, the optimizer simultaneously tries to minimize the loss with respect to the neural network trainable parameters while maximizing the loss with respect to the loss terms to ensure that the loss function terms are properly scaled. 
\end{itemize}

As discussed in \S~\ref{sec: SFNE description}, artificial neural networks in smooth function approximators face a challenge when being applied to problems with sharp transitions in the output. While this challenge was briefly described in \S~\ref{sec: SFNE description}, we further address the strategies implemented in this study to overcome this challenge before presenting the training results.

\begin{itemize}
    \item \textit{Activation function:} As we discussed in \S~\ref{sec: beam SFNE architecture design}, we utilized the Snake activation function in the output layers of SFNEs (in $\text{NN}_{Out}$) for the beam problem. This function, proposed by Ziyin \etal~\cite{ziyin2020neural}, is designed to help the network learn high-frequency components of the output signal. The Snake function is a periodic function that can capture oscillatory behavior better than traditional activation functions like ReLU or sigmoid. Note that activation functions based on harmonic functions ($sin(x), cos(x), sin^2(x)$ or $cos^2(x)$) as activation result in several local minima in the training loss, which prevents training convergence. This issue is addressed in the snake activation function by the trainable scaling factor $a_1$ (see Eq.~\ref{eq: revised-activation}), which allows the function to adapt to the specific characteristics of the data without adding several local minima to the loss function. A thorough discussion and comparison of this activation function with other activation functions can be found in~\cite{ziyin2020neural}.  
    \item \textit{Deep MLPs:} The proposed SFNE architecture employs deep MLPs. The depth of these networks significantly impacts their ability to learn complex patterns and sharp variations. Recall that deep networks have higher expressivity compared to shallow networks, meaning they can approximate a wider range of functions, which is particularly useful for capturing sharp variations in the data.
    \item \textit{Bidirectional recurrent neural networks:} BRNNs were chosen over RNNs as they propagate information in both forward and backward directions, which helps capture the dynamics of the system more accurately (see also the discussions in \S~\ref{sec: SFNE description}).
    \item \textit{Loss function and learning rate adjustment:} A revised loss function was used for the beam problem, consisting of a weighted summation of MSE and normalized MSE (Eq.~\ref{eq: revised loss}-\ref{eq: revised training}). We also used a rate scheduling method to adjust the learning rate (\textit{StepLR} for rod and \textit{ReduceLROnPlateau} for beams), resulting in a dynamic adjustment of the learning rate. These adjustments prevent the network from overfitting on training data and enable accurately capturing the transient response of the structure. 
    \item \textit{Ensemble methods:} We used an ensemble method (bagging) to better capture abrupt changes in the data. Each model in the ensemble may specialize in different aspects associated with the data, potentially allowing sharp transitions to be better represented by the bag of models.  
\end{itemize}

\begin{figure}[!h]
	\centering
	\includegraphics[width=\linewidth]{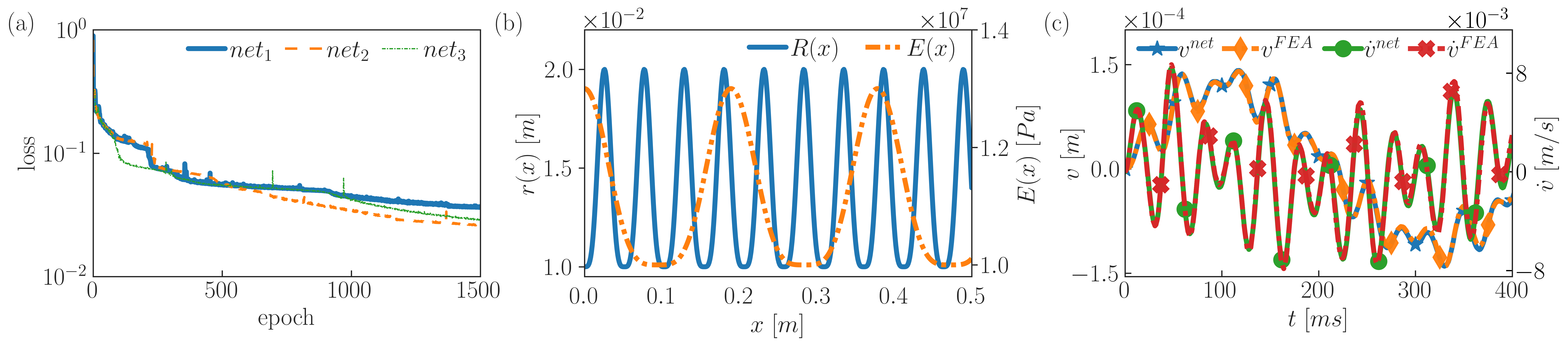}
	\caption{(a) Trend of the training loss function for the ensemble of three inhomogeneous beam network elements. (b) Distribution of Young's modulus $E(x)$ and cross-section radius $R(x)$ for $beam_1$. (c) Network elements predictions for $beam_1$ compared with the FE solution ($\Box^{FEA}$). The results correspond to the point $x=0.44~m$. $v^{net}$ and $\dot{v}^{net}$ are obtained by averaging the predictions of the three networks within the ensemble.}\label{fig: beam_training_and_sample}
\end{figure}

We trained three network elements for the simulation of inhomogeneous beams subject to uniformly distributed load and zero initial conditions. Figure~\ref{fig: beam_training_and_sample}a shows the trend of the loss function for the network elements. The average prediction error ($e_r$) of the beam network element ensemble for the training and test datasets is $0.78~\%$ and $0.81~\%$, respectively. Note again that the error is calculated on the basis of the average prediction of the network models within the ensemble.

Figure~\ref{fig: beam_training_and_sample}c shows the response of a sample (inhomogeneous) beam, labeled $beam_1$. The distributions of $E$ and $R$ for $beam_1$ are plotted in Figure~\ref{fig: beam_training_and_sample}b. In $beam_1$, the input load excitation frequency is $\omega_0 =377.0~rad/s$. The relative prediction error for $beam_1$ is $0.55~\%$. Comparing the network ensemble predictions with the FE-based solution shows that the network accurately predicted the response. 

\subsection{Case~5: Transient response of inhomogeneous beams to arbitrary initial conditions}\label{sec: IV beam}
In this section, we extend the network elements developed in \S~\ref{sec: Zero_init_beam} to nonzero initial conditions ($v_0(x)$ and $\dot{v}_0(x)$) for the beam configuration shown in Figure~\ref{fig: beams_schematic}. We follow a similar procedure as in \S~\ref{sec: initial BV rod} to incorporate the initial conditions in the inputs of the network elements. 
The general steps for the development of SFNEs are similar to \S~\ref{sec: initial BV rod}; here below we briefly summarize the salient details.

\begin{figure}[!h]
	\centering
	\includegraphics[width=.8\linewidth]{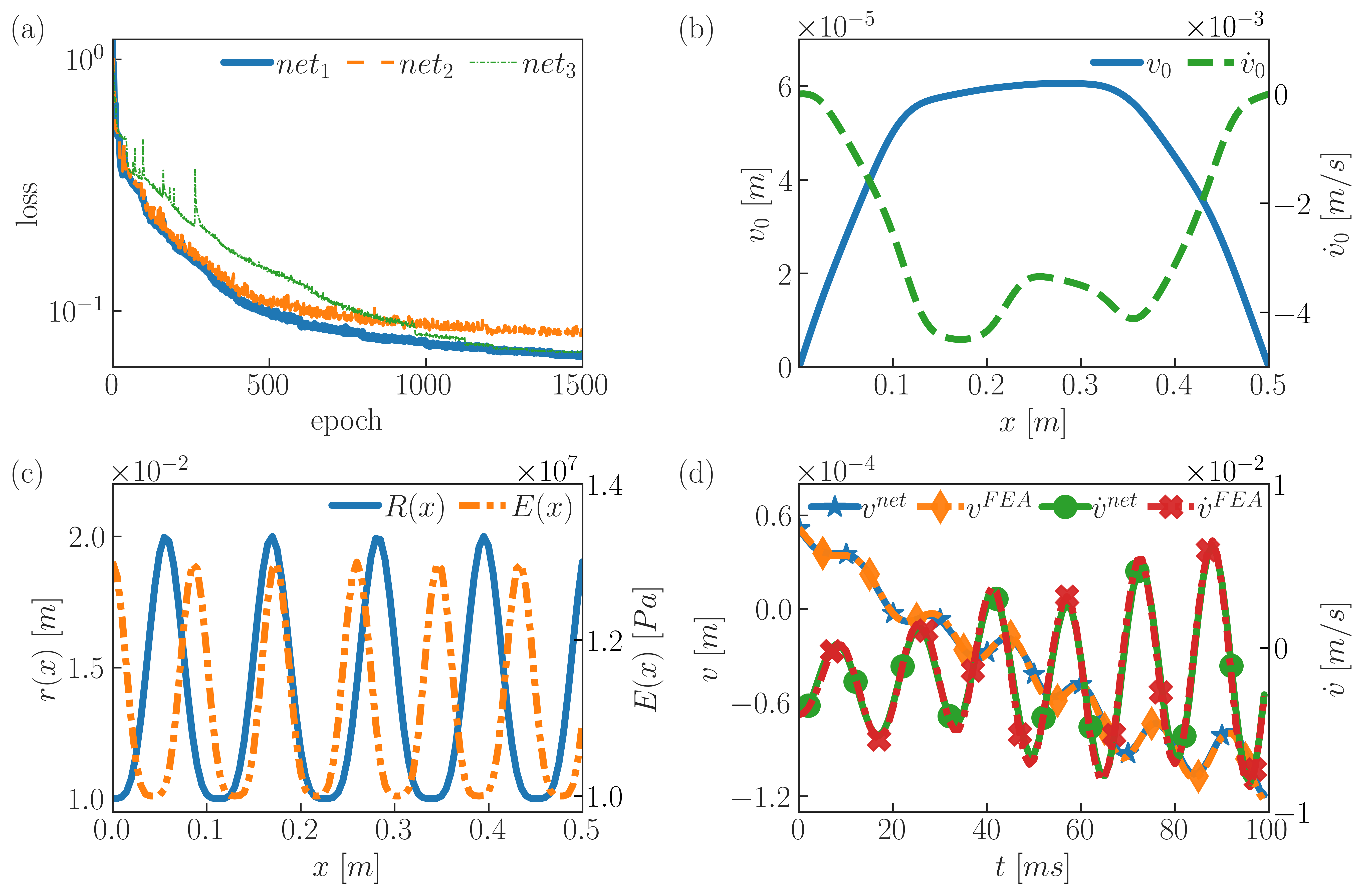}
	\caption{(a) Trend of training loss function for the ensemble of the beam network elements for nonzero initial conditions. (b) Initial conditions of the sample problem $beam_2$. (c) Distribution of Young's modulus $E(x)$ and cross-section radius $R(x)$ for $beam_2$. (d) Transverse deformation and velocity of $beam_2$. The results are reported for $x=0.25~m$ (arbitrarily selected) and $\omega_0= 428.18~rad/s$.}\label{fig: Initial beam_training_and_sample}
\end{figure}

The training dataset was generated based on the dataset discussed in \S~\ref{sec: Zero_init_beam}. We followed the procedure proposed in \S~\ref{sec: initial value rod training} to generate data samples with nonzero initial conditions from the data sample of \S~\ref{sec: Zero_init_beam}. Specifically, we randomly selected an initial time step $0<t_s<300~ms$, saved $v(x, t_s), \dot{v}(x,t_s)$ as initial conditions ($v_0(x)$ and $\dot{v}_0(x)$), and extracted the beam response at the following 100 time steps and formed a data sample with nonzero initial conditions. The network architecture used for the problem is the same as in the previous section (see Tables~\ref{table: Beam_netArch}~and~\ref{tab: CNN1_net_architec}), except for one adjustment. We revised the number of input channels of $\text{CNN}_S$ to 4 to include the initial conditions in the static input channel of the network element. Hence, $In^S$ is an array with four rows defined by $\{E(x), R(x), v_0(x), \dot{v}_0(x)\}$. 
We used the same training scheme as in \S~\ref{sec: training_zero_init_beam_network} and trained an ensemble of three network elements. Figure~\ref{fig: Initial beam_training_and_sample}a shows the network elements training loss versus training epoch. The average train and test relative prediction errors of the network ensemble are $1.22~\%$ and $1.31~\%$, respectively, showing excellent prediction accuracy of the ensemble. The results also demonstrate the network elements' capability in the simulation of problems with nonuniform properties. 

Similarly to the previous sections, to showcase the ensemble performance, we also present a sample test problem labeled $beam_2$. Note that this problem is randomly selected from the test dataset. The initial conditions and properties ($E(x)$ and $R(x)$) of $beam_2$ are presented in Figure~\ref{fig: Initial beam_training_and_sample}b~and~c, respectively. The beam is subject to uniformly harmonic distributed load with $\omega_0= 428.18~rad/s$. $beam_2$ deformation is depicted in Figure~\ref{fig: Initial beam_training_and_sample}d. For $beam_2$, the average error is $e_r=1.06\%$.  The results show again that the ensemble calculated the beam deformation with a very small error when compared to the FE solution.  

A few computational aspects are connected to the development of the network elements. For the beam problem discussed in this section, there is a significant increase in the computational cost of the training dataset generation, because a FE software must be used to generate the training data. We used the cluster node described in \S~\ref{sec: computational costs} to build the training data. The average simulation time of each sample was $33.4~s$. Note again that although building the training data is resource exhaustive, the training data must be generated only once.
In addition, these databases are general and could be used for any future modeling need that requires the specified elements type (either rods or beams, for the case of this study). Therefore, similarly to the case for many databases available for the most diverse applications, such as the ImageNet dataset that is used for computer vision~\cite{deng2009imagenet} or Google Books Ngrams dataset used for natural language processing~\cite{lin2012syntactic}, these databases of structural elements can also become standardized and available to the scientific community for use.

\begin{figure}[!h]
	\centering
	\includegraphics[width=.45\linewidth]{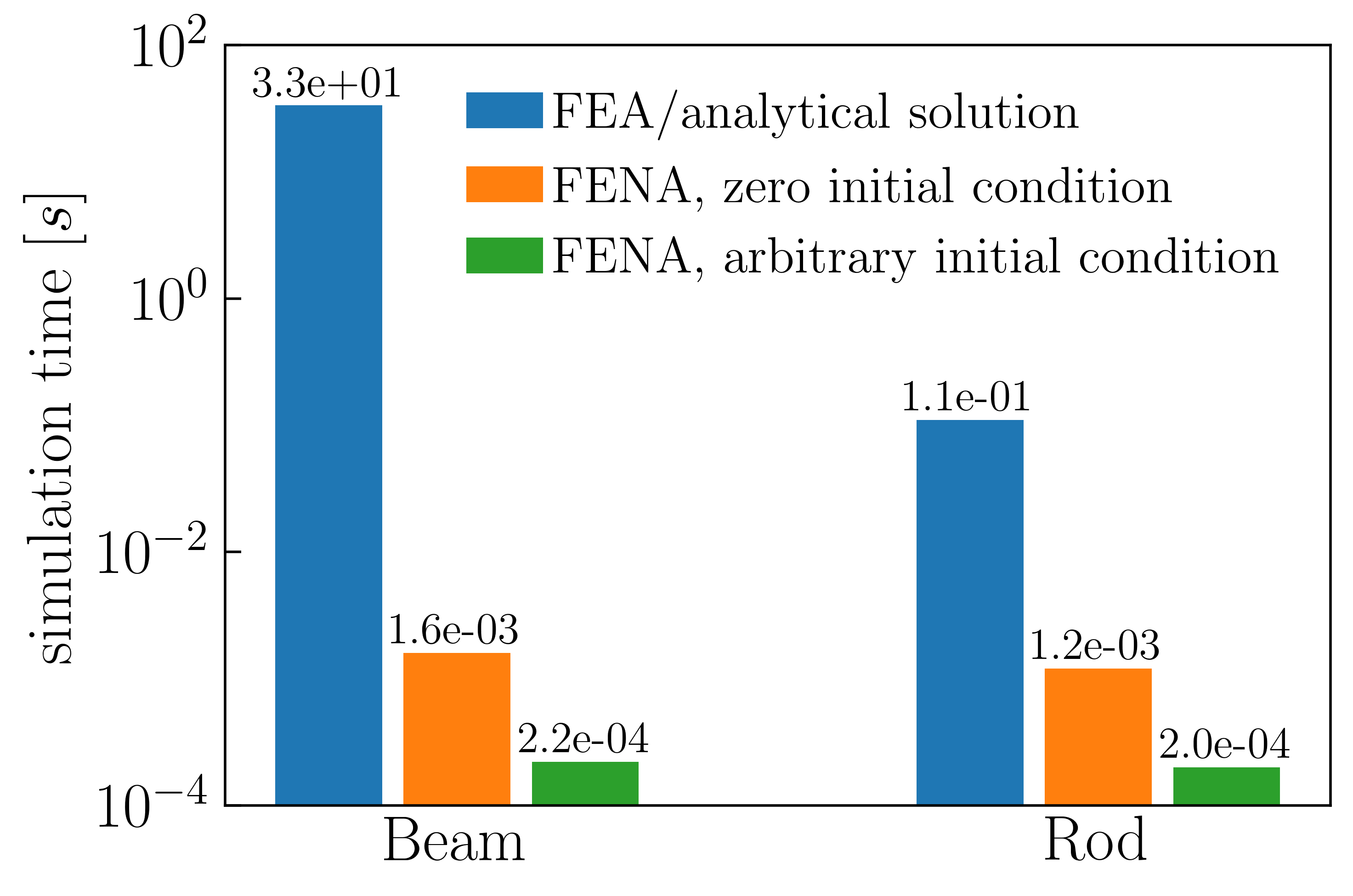}
	\caption{Average computational time of FENA compared to FEA and analytical solutions for the transient simulation of beams (\S~\ref{sec: beam}) and rods (\S~\ref{sec: rod}).}\label{fig: sim time}
\end{figure}

The average training time of the network elements in Case~5 was $4.3~hr$, and for Case~4 the training time averaged $8.9~hr$ (see \S~\ref{sec: Training_BV_Problem} for the training hardware details). The training time of the beam network elements decreased compared to the rod network elements because the training dataset size was smaller for the beam; hence, each training epoch was performed in a shorter period of time. 
On the other hand, as seen in Figure~\ref{fig: sim time}, despite the increase in the complexity of the beam (compared to the cases in \S~\ref{sec: rod}), the computational cost of network elements execution has slightly increased. Specifically, the SFNEs developed for zero initial conditions require an average of $0.0016~s$ to predict the response, and the SFNEs for nonzero initial conditions take $0.00022~s$ (compare with $0.0012~s$ and $0.0002$ for the rod SFNEs). Figure~\ref{fig: sim time} compares the average solution time of FENA to the beam FEA and rod analytical solutions.
Note again that these numbers are calculated based on the average prediction time of the test dataset. Results show that the SFNEs computational time does not scale with the complexity of the physics of the problem being simulated. On the contrary, the computational time of the FE-based solution rapidly grows with the complexity of the governing physics and conditions (see Figure~\ref{fig: sim time}). Comparing the computational costs (FEA-based and FENA) demonstrates the potential of the proposed method when applied to the simulations of complex physical systems. 

\section{Conclusions}

The concept of finite element network analysis (FENA) was extended to the transient analysis of structural elements. Two key novel concepts were introduced to make this extension possible: 1) the super finite network element (SFNE), and 2) the network concatenation in time. The former is a fundamental neural network architecture specifically designed to address the needs of transient simulations of physical systems, while the latter allows extending the computational window of the network element far beyond the maximum number of time steps the network element was trained for.

The SFNE leverages the unique properties of bidirectional recurrent neural networks (BRNN) that provide the foundation for building network elements for the transient simulation of structures. The core BRNN network block within the SFNEs enables learning of the transient behavior of physical systems. The SFNE is also equipped with supplementary network blocks that allow the incorporation of various types of input loads, boundary and initial conditions, and material and geometric properties, further enhancing its versatility and adaptability for different types of structural simulations.
SFNEs can be specialized for a specific class of problems based on the relevant properties, applied external loads, and boundary conditions, thereby enabling them to serve as problem solvers for that particular class of problems.
This remarkable property separates FENA from other network-based simulation approaches because, unlike other neural network-based methods, FENA network elements do not require \textit{ad hoc} training for any minor change in the problem conditions.

The concept of network concatenation was also extended to enable the simulation of transient problems. Concatenation enables a virtually unbounded simulation time window, by allowing the use of network elements that were trained on data defined over limited time frames. 
In reality, the number of consecutive concatenations, hence the total time window for computations, cannot be infinite due to some level of error accumulation that is still present. Nevertheless, the concatenation approach provides a drastic augmentation of the computational window (at least an order of magnitude larger than the initial training window) with respect to the one used during the training phase. This is a critical property when taking into account the significant computational cost associated with the numerical generation of data to train the network for transient simulations. 

To showcase the capabilities of the proposed concept, we developed a library of network elements for two types of one-dimensional structural elements, and including different types of initial and boundary conditions. All network elements were built on the basis of the proposed SFNE fundamental network architecture. 
Several test cases were presented to validate the approach and assess the prediction performance of FENA. Simulation tests showed excellent predictive capabilities of FENA, with prediction errors mostly below $1 \% $ and significant reductions in computational time (by at least 2 orders of magnitude) compared to traditional numerical (finite element) solution approaches. 

Although the examples presented in the present study are dedicated to transient simulations in structures, the methodologies introduced are general and can be extended to other physical systems. It can be envisioned that extending FENA to more complex and nonlinear systems would further demonstrate its computational benefits over traditional techniques.

\section{Acknowledgments}
This material is based upon work supported by, or in part by, the U. S. Army Research Office under contract number W911NF2410076 and the National Science Foundation under grant number CMMI 2330957.

\bibliographystyle{unsrt}
\bibliography{Refs}

\end{document}